\documentclass[11pt,a4paper]{amsart}
\usepackage[T1]{fontenc}
\usepackage[margin=3cm]{geometry}

\usepackage{amsmath,amsfonts,amssymb,amscd,amsthm, mathtools}
\usepackage[usenames,dvipsnames,svgnames]{xcolor}

\usepackage[utf8]{inputenc}
\usepackage{graphicx}
\usepackage{caption}
\usepackage{subcaption}
\usepackage{doi}
\usepackage{autonum}  
\usepackage{makecell}

\usepackage{hyperref} 
\usepackage{epstopdf}

\usepackage{acronym}
\usepackage{enumitem}
\usepackage{multirow}
\usepackage{booktabs}
\usepackage[norelsize]{algorithm2e} 
 
\usepackage{tikz}
\definecolor{myellow}{RGB}{255,230,128}
\definecolor{gray20}{RGB}{204,204,204}
\definecolor{mygray}{RGB}{204,204,204}
\definecolor{mygreen}{RGB}{138,203,95}
\definecolor{myblue}{RGB}{77,151,214}

\usepackage[textsize=small]{todonotes}

\acrodef{cg}[CG]{conjugate gradient}
\acrodef{vef}[vef]{vertex, edge, and face}
\acrodefplural{vef}[vefs]{vertices, edges, and faces}
\acrodef{glob}[glob]{global object}
\acrodefplural{glob}[globs]{global objects}
\acrodef{fe}[FE]{finite element}
\acrodefplural{fe}[FEs]{finite element}
\acrodef{pb}[PB]{physics-based}
\acrodef{rpb}[rPB]{ relaxed \ac{pb} }
\acrodef{dof}[DOF]{degree of freedom}
\acrodefplural{dof}[DOFs]{degrees of freedom}
\acrodef{dg}[DG]{discontinuous Galerkin}
\acrodef{vms}[VMS]{variational multiscale}
\acrodef{sps}[SPS]{symmetric projection stabilization}
\acrodef{agfem}[agFEM]{aggregated finite element method}
\acrodef{xfem}[XFEM]{extended finite element method}
\acrodef{agfe}[agFE]{aggregated finite element}
\acrodef{pde}[PDE]{partial differential equation}
\acrodef{bddc}[BDDC]{balancing domain decomposition by constraints}
\acrodef{dd}[DD]{domain decomposition}
\acrodef{pbbddc}[PB-BDDC]{physics-based \ac{bddc}}
\acrodef{rpbbddc}[rPB-BDDC]{relaxed \ac{pbbddc}}
\acrodef{hts}[HTS]{High Temperature Superconductors}

\newtheorem{theorem}{Theorem}[section]

\newtheorem{proposition}[theorem]{Proposition}

\newtheorem{remark}[theorem]{Remark}



\def\REV#1{#1}

\def\sect#1{Sect.~\ref{#1}}
\def\eq#1{Eq.~(\ref{#1})}
\def\fig#1{Fig.~\ref{#1}}
\def\tab#1{Tab.~\ref{#1}}

\def\FEMPAR{{\texttt{FEMPAR}}~}

\def\n{{\boldsymbol{n}}}

\def\grad{{\boldsymbol{\nabla}}} 
\def\curl{\grad \times }
\def\norm#1{\|#1\|}
\def\bs#1{\boldsymbol{#1}}
\def\massc{\beta}
\def\curlc{\alpha}

\def\uc{\bs{u}}

\def\w{\bs{w}}
\def\vc{\bs{v}}
\def\f{\bs{f}}
\def\t{\bs{t}}
\def\n{\bs{n}}

\def\Hcurl{H({\rm \curl };\Omega)}               
\def\Hcurlo{H_0({\rm \curl };\Omega)}            
\def\Hcurlti{H({\rm \curl };\tilde{\Omega}_i)}   
\def\lagsh{\phi}                                 
\def\nedsh{\boldsymbol{\varphi}}                 
\def\Xbddc{{\widetilde {X}}_h}                   
\def\Xg{\REV{\widehat{X}_h}}                     
\def\X{\REV{X_h}}                                
\def\Xi{{X}_h^i}                                 
\def\Xj{{X}_h^j}                                 
\def\V{\REV{\widehat{V}_h}}                      
\def\Vi{{V_{h}^i}}                               

\def\Cb{\REV{ \widehat{\mathcal{Q}}}}            
\def\Cbinv{\REV{\widehat{\mathcal{Q}}^{-1}}}     
\def\Cbwinv{\REV{{\mathcal{{Q}}}^{-1}}}          
\def\Cbwtinv{\REV{{\mathcal{{Q}}}^{-T}}}         
\def\Cbw{\REV{{\mathcal{Q}}}}                    
\def\Cbi{{\mathcal{{Q}}}_i}                      
\def\Cbj{{\mathcal{{Q}}}_j}                      
\def\Cbiinv{\mathcal{{Q}}^{-1}_i}                

\def\Abddc{{\widetilde {\mathcal{A}}_h }}        
\def\Ad{\REV{{\mathcal{A}}_h}}                   
\def\Ai{{\mathcal{A}}_{h}^{i}}                   
\def\Ag{\REV{\widehat{\mathcal{A}}_{h}}}         
\def\Ab{\REV{\widehat{\mathcal{A}}_{h,0}}}       
\def\Abinv{\REV{\widehat{\mathcal{A}}^{-1}_{h,0}}} 

\def\weiope{{ {\mathcal{W}_h}}}
\def\preope{{ {\mathcal{P}}}}

\def\tri{{\mathcal{T}_h}}

\def\partition{{ {\Theta}}}
\def\pbpartition{{ {\Theta}_{\rm pb}}}
\def\rpbpartition{{ {\Theta}^r_{\rm pb}}}

\def\neigho{{ {\rm neigh}_{\partition}}}
\def\neigh{{ {\rm neigh}_{\pbpartition}}}
\def\verts{{ {\mathcal{V}}}}
\def\edgets{{ \REV{\mathcal{S}}}}
\def\facets{{ {\mathcal{F}}}}

\begin{document}

\title[Scalable solvers for complex electromagnetics problems]{Scalable solvers for complex electromagnetics problems}

\author[S. Badia]{Santiago Badia$^{\dag,\ddag}$}

\author[A. F. Mart\'in]{Alberto F. Mart\'in$^{\dag,\ddag}$}

\author[M. Olm]{Marc Olm$^\dag$}

\thanks{$\ddag$ Centre Internacional de M\`etodes Num\`erics en Enginyeria, Esteve Terrades 5, E-08860 Castelldefels, Spain. $\dag$ Universitat Polit\`ecnica de Catalunya, Jordi Girona1-3, Edifici C1, E-08034 Barcelona, Spain. \\ SB gratefully acknowledges the support received from the Catalan Government through the ICREA Acad\`emia Research Program. MO gratefully acknowledges the support received from the Catalan Government through the FI-AGAUR grant. This work has been partially funded by the project MTM2014-60713-P from the ``Ministerio de Econom\'ia, industria y Competitividad'' of Spain. The authors thankfully acknowledge the computer resources at Marenostrum-IV and the technical support provided by the Barcelona Supercomputing Center (RES-ActivityID: FI-2018-3-0029). E-mails: {\tt sbadia@cimne.upc.edu} (SB), {\tt amartin@cimne.upc.edu} (AM), and {\tt molm@cimne.upc.edu} (MO)}

\date{\today}

\begin{abstract}

In this work, we present scalable balancing domain decomposition by constraints methods for linear systems arising from arbitrary order edge finite element discretizations of multi-material and heterogeneous 3D problems.  In order to enforce the continuity across subdomains of the method, we use a partition of the interface objects (edges and faces) into sub-objects determined by the variation of the physical coefficients of the problem.  For multi-material problems, a constant coefficient condition is enough to define this sub-partition of the objects. For arbitrarily heterogeneous problems, a relaxed version of the method is defined, where we only require that the maximal contrast of the physical coefficient in each object is smaller than a predefined threshold.  Besides, the addition of perturbation terms to the preconditioner is empirically shown to be effective in order to deal with the case where the two coefficients of the model problem jump simultaneously across the interface.  The new method, in contrast to existing approaches for problems in curl-conforming spaces \REV{does not require spectral information} whilst providing robustness with regard to coefficient jumps and heterogeneous materials.   A detailed set of numerical experiments, which includes the application of the preconditioner to 3D realistic cases, shows excellent weak scalability properties of the implementation of the proposed algorithms.  
\end{abstract}

\maketitle




\section{Introduction}\label{sec:int}

Realistic simulations in electromagnetic problems often involve multiple materials (e.g., dielectric and conducting materials), which may imply high contrasts in the coefficients describing the physical properties of the different materials. Besides, the behaviour of conducting materials may be modelled by highly variable, heterogeneous coefficients. This problem definition inevitably leads to high condition numbers for the resulting linear systems arising from curl-conforming \ac{fe} discretizations of the corresponding \ac{pde}, which pose great \REV{challenges} for solvers. Furthermore, the design of solvers for $H$(curl)-conforming approximations poses additional difficulties, since the kernel of the curl operator is non-trivial. Consequently, for realistic electromagnetic simulations in 3D, the use of robust iterative solvers is imperative in terms of complexity and scalability. In this work, we will focus on the development of robust \ac{bddc} preconditioners for problems posed in $H$(curl) involving high variation of the coefficients for the corresponding \ac{pde}.

\ac{bddc} preconditioners \cite{dohrmann_preconditioner_2003} belong to the family of non-overlapping \ac{dd} methods \cite{DD_toselli}. They can be understood as an evolution of the earlier Balancing \ac{dd} method \cite{mandel_bdd}. These methods rely on the definition of a \ac{fe} space with relaxed inter-element continuity, which is defined by choosing some quantities to be continuous across subdomain interfaces, i.e., the \emph{coarse} or \emph{primal} \acp{dof}. Then, the continuity of the solution at the interface between subdomains is restored with an averaging operator. The method has two properties that make it an outstanding candidate for extreme scale computing, namely it allows for aggressive coarsening and computations among the different levels can be performed in parallel. Outstanding scalability results have been achieved by an implementation in the scientific computing software \FEMPAR \cite{fempar-web-page,badia-fempar}, which exploits these two properties in up to almost half a million cores and two million subdomains (MPI tasks) \cite{badia_multilevel_2016}. Another work showing excellent scalability properties up to two hundred thousand cores is \cite{zampini_PETSc_scalability}, which is implemented in the software project PETSc \cite{petsc-web-page}.

The main purpose of this work is to construct \ac{bddc} methods for the linear systems arising from arbitrary order edge (N\'ed\'elec) \ac{fe} discretizations of heterogeneous electromagnetic problems. An analysis for 3D FETI-DP\footnote{FETI-DP algorithms \cite{FETIDP_farhat} are closely related to \ac{bddc} methods. In fact, it can be shown that the eigenvalues of the preconditioned operators associated with \ac{bddc} and FETI-DP are almost identical \cite{mandel_algebraic_2005, li_feti-dp_2006, brenner_bddc_2007}.}
  algorithms with the lowest order N\'ed\'elec elements of the first kind was given by Toselli in \cite{toselli_dual-primal_2006}, who argued that the difficulty of iterative substructuring methods for edge element approximations mainly lies in the strong coupling between the energy of subdomain faces and edges. In short, no efficient and robust iterative substructuring strategy is possible with the standard basis of shape functions for the edge \ac{fe} (see \cite{molm_nedelec}). A suitable change of basis was introduced in \cite{toselli_dual-primal_2006} for lowest order edge elements and box-subdomains. Besides, an extension to arbitrary order edge \acp{fe} and subdomain geometrical shapes is presented in \cite{Zampini2017}. 
 In this work, we will offer some new insights in the definition and construction of the change of basis for the latter general case. As pointed out in \cite{dohrmann_recent_tools}, the change of variables can be implemented in practice with just a few simple modifications to the standard \ac{bddc} algorithm \cite{dohrmann_preconditioner_2003}. 

Modern \ac{bddc} methods \cite{Pechstein2017} propose coarse space enrichment techniques that adapt to the variation of coefficients of the problem \cite{mandel_adaptive_2007, sousedik_adaptive-multilevel_2013, calvo_adaptive, spillane_2011, spillane_2013, klawonn_adaptive_2016}, where coarse \acp{dof} are adaptively selected by solving generalized eigenvalue problems. This approach is backed up by rigorous mathematical theory and has been numerically shown to be robust for general heterogeneous problems.
On the other hand, several different scalings have been proposed for the averaging operator in the literature to improve the lack of robustness of the \emph{cardinality} (i.e., arithmetic mean) scaling for coefficient jumps. The \emph{stiffness}\footnote{weighted averages with the diagonal entries of the operator for every \ac{dof}} scaling takes more information into account but can lead to poor preconditioner performance with mildly varying coefficients  \cite{Pechstein2011}. 
\REV{A robust approach} is the \emph{deluxe} scaling, first introduced in \cite{dohrmann_recent_tools} for 3D problems in curl-conforming spaces. It is based on the solution of local auxiliary
Dirichlet problems to compute efficient averaging operators \cite{dohrmann_bddc_2016, Pechstein2017, duk_soon_bddc, zampini_PETSc_scalability, zampini_adaptive, zampini_flow}, involving dense matrices per subdomain vertex/edge/face. 
However, to solve eigenvalue and auxiliary problems is expensive and extra implementation effort is
required as coarse spaces in \ac{dd} methods are not naturally formulated as eigenfunctions. 

The main motivation of this paper is to construct robust \ac{bddc} preconditioners for problems in curl-conforming spaces that keep the simplicity of the standard \ac{bddc} method, i.e., to avoid the spectral solvers of adaptive versions, whereas keeping robustness and low computational cost. In order to do so, we follow the idea of the \ac{pbbddc} preconditioner, presented in \cite{BadiaNguyen_physics_based} for problems in grad-conforming spaces.

Based on the fact that \ac{bddc} methods ( and \ac{dd} methods in general ) are robust with regard to jumps in the material coefficients when these jumps are aligned with the partition \cite{DD_toselli, Klawonn_heterogeneous}, one can use a \ac{pb}-partition obtained by aggregating elements of the same (or similar) coefficient value. However, using this type of partition can lead to a poor load balancing among subdomains and large interfaces. To overcome this situation, the \ac{pbbddc} respects the original partition (well-balanced) but considers a sub-partition of every subdomain based on the physical
coefficients, leading to a partition of the objects into sub-objects defined according to the variation of the coefficients. Consequently, the method is also based on an enrichment of the coarse space but with the great advantage of not requiring to solve eigenvalue or auxiliary problems, i.e., the simplicity of the original \ac{bddc} preconditioner is maintained. 
\REV{ On the other hand, the \ac{pbbddc} methods involve a richer
  interface with the application software, e.g., access to the
  physical properties of the problem, which is in line with the
  philosophy of the \texttt{FEMPAR} library, i.e., a tight interaction
  of discretization and linear solver steps to fully exploit the
  mathematical structure of the PDE operator.}
The \ac{pbbddc} preconditioner turned out to be one order of magnitude faster than the \ac{bddc} method with deluxe scaling in \cite{zampini_PETSc_scalability} for linear elasticity and thermal conductivity problems with high contrast.

Our problem formulation arises from the time-domain quasi-static approximation to the Maxwell's Equations for the magnetic field (see \cite{molm_hts}), which involves two different operators, the mass and double curl terms. This fact certainly poses more complexities than the ones faced in \cite{BadiaNguyen_physics_based} for the \ac{pbbddc} solver, since it has to deal with the interplay of both (simultaneous) coefficient jumps. Our solution is to propose a simple technique to recover the scenario where only one coefficient has a jump across interfaces: we will add a perturbation at the preconditioner level so that the perturbed formulation does not involve a jump for the mass-matrix terms across interfaces. The effectiveness of the technique will be empirically shown. 
In order to extend the \ac{pbbddc} algorithm to heterogeneous materials, a relaxed definition of the \ac{pb}-partition will be stated where we only require that the maximal contrast of the physical coefficient in each
\ac{pb}-subdomain is smaller than a predefined threshold. The threshold can be chosen
so that the condition number is reasonably small while the size of the coarse problem
is not \emph{too large}.

The article outline is as follows. The problem is defined in \sect{sec-psetting}, where basic definitions are introduced. \sect{sec-pbbddc} is devoted to the presentation of the \ac{pbbddc} preconditioner for heterogeneous 3D problems in $H$(curl). In \sect{sec-implementation} we will give some implementation insights, based on our experience through the implementation of the algorithms in the scientific software project \FEMPAR. In \sect{sec-results}, we present a detailed set of numerical experiments, covering a wide range of cases and applications for the \ac{pbbddc} preconditioner. Finally, some conclusions are drawn in \sect{sec-conclusions}.  

\section{Problem setting}\label{sec-psetting}

Let us consider the boundary value Maxwell problem on a physical domain
$\Omega \subset \mathbb{R}^3$:
\begin{align}
   \curl(\curlc \curl \uc) + \massc \uc = \f \quad & \hbox{in} \, \Omega, \label{eq-strong_maxwell} \\
   \n \times ( \uc \times \n ) = 0 \quad & \hbox{on} \, \partial \Omega , 
\end{align}
where $\curlc \geq 0$, $\massc>0$ are the resistivity and the magnetic permeability of the materials, respectively, $\n$ is a unit normal to the boundary and $\curl$ is the 3D curl operator (see \cite{monk_finite_2003}). For the sake of simplicity, we consider homogeneous Dirichlet conditions, i.e., zero
tangential traces on $\partial\Omega$. Nevertheless, all the developments in this work can readily be applied to Neumann and/or inhomogeneous conditions, see \cite{molm_nedelec} for proper definitions. In order to pose the weak form of the problem, let us define the functional space 
\begin{align}
  \Hcurl \doteq \{ \vc \in L^2(\Omega)^3 \, : \, \curl \vc \in L^2(\Omega)^3 \}, 
\end{align}
and its subspace that satisfies homogeneous Dirichlet boundary conditions,
\begin{align}
  \Hcurlo \doteq \{ \vc \in \Hcurl \, : \, \n \times (\uc \times \n) = 0 \, \hbox{on} \, \partial \Omega \}.
\end{align} 
Besides, we will also make use of the space 
\begin{align}
  H^1(\Omega) \doteq \{ v \in L^2(\Omega) \, : \, \grad v \in L^2(\Omega)^3 \}.
\end{align} 
Functions in $\Hcurl$ are approximated by edge \ac{fe} methods of arbitrary order, which we represent by $\Xg \subset \Hcurl$. In addition, functions in $H^1(\Omega)$ are approximated by standard scalar, continuous Lagrangian \ac{fe} methods, which we represent by $\V \subset H^1(\Omega)$. 
The weak form of the boundary value Maxwell problem in \eq{eq-strong_maxwell} reads: find $\uc \in \Hcurlo$ such that  
\begin{align} 
\Ag ( \uc, \vc ) = ( \f, \vc ), \qquad \forall \vc \in \Hcurlo, 
\end{align} 
where 
\begin{align}\label{eq-operator}
\Ag (\uc,\vc) =  \int_\Omega \left[ (\curlc \curl \uc)\cdot(\curl \vc) + \massc \uc \cdot \vc \right] dx, \qquad (\f,\vc) = \int_\Omega \f \cdot \vc dx.
\end{align} 

\subsection{Domain partition} 
Let us consider a bounded polyhedral domain $\Omega \subset \mathbb{R}^3$. Let $\tri$ be a partition of $\Omega$ into a set of tetrahedral or hexahedral cells $K$. For every cell $K\in\tri$ consider its set of vertices $\verts_K$, edges $\edgets_K$, or faces $\facets_K$. They constitute the set of geometrical entities of the cell (excluding itself) as $\mathcal{G}_K = \verts_K \cup \edgets_K \cup \facets_K$. The union of these sets for all cells is represented with $\mathcal{G} \doteq \cup_{K\in\tri} \mathcal{G}_K$. 
We consider a partition $\partition$ of the domain $\Omega$ into non-overlapping subdomains $\tilde
\Omega_i$, $i=1,\ldots,\tilde N$ obtained by aggregation of elements $K\in\mathcal{T}_h$. These subdomains are assumed to be such that
the computational cost of solving the discrete Maxwell problem in the different
subdomains leads to a well-balanced distribution of computational loads among processors in memory distributed platforms. We denote by $\Gamma(\partition)$ the interface of the partition $\partition$, i.e., $\Gamma(\partition) = \cup_{\tilde\Omega_i\subset\Omega} ~\partial \tilde\Omega_i \setminus \partial\Omega$.
Every subdomain $\tilde \Omega_i \subset \Omega$ can be also partitioned
into the smallest set of subdomains $\Omega_{ij}$, $j=1,\ldots,N_i$, such that the
material properties $(\curlc,\massc)$ in \eq{eq-strong_maxwell} are constant at every $\Omega_{ij}$. For obvious reasons, we call this sub-partition a \ac{pb}-partition and will be denoted by $\pbpartition$. Clearly, the resulting global $\pbpartition$ is also a partition of $\Omega$, and there is a unique $\mathcal{D} \in \partition$ for every $\mathcal{D'} \in \pbpartition$ such that $\mathcal{D'} \subset \mathcal{D}$. We consider a global numbering for the \ac{pb}-subdomains, i.e., $\Omega_{k}$, $k=1,\ldots,N$,
having a one-to-one mapping between the two indices labels. Analogously, we define the interface of the \ac{pb}-partition as $\Gamma(\pbpartition) = \cup_{\Omega_k\subset\Omega} ~\partial\Omega_k \setminus \partial\Omega$.

\subsection{Finite Element spaces}
Let us define the \ac{fe} spaces $\Xi \doteq \Xg(\tilde \Omega_i) \subset \Hcurlti$ for every subdomain $\mathcal{D}\in\partition$, and the
corresponding Cartesian product space $\X = \Pi_{i=1}^{\tilde N} \Xi$. Note that functions belonging to this space are allowed to have discontinuous tangent traces across the interface $\Gamma(\partition)$. The global space in which the global problem is sought, i.e., $\Xg$, can be understood as the subspace of functions in $\X$ that have continuous tangent traces across $\Gamma(\partition)$. We can now define the subdomain \ac{fe} operator $\Ai: \Xi \rightarrow {\Xi}'$, $i=1,\ldots,\tilde N$, as $\Ai( \uc_i, \vc_i) = \int_{\tilde \Omega_i} \left[ (\curlc \curl \uc_i)\cdot (\curl \vc_i) + \massc \uc_i \cdot \vc_i \right] dx$ for all $\uc_i,\vc_i \in \Xi$. Then, the sub-assembled operator $\Ad: \X \rightarrow {\X}'$ is defined as $\Ad(\uc, \vc) = \prod_{i}^{\tilde N} \Ai( \uc_i, \vc_i)$, in which contributions between subdomains have not been assembled. The assembled operator $\Ag: \Xg \rightarrow {\Xg}'$ (see \eq{eq-operator}) is the Galerkin projection of the operator $\Ad$ onto $\Xg$.

The space of edge \ac{fe} functions can be represented as the range of an interpolation operator $\pi^h$, which is well-defined for sufficiently smooth functions $\uc \in \Hcurl$, by 
\begin{align}\label{eq-int_op}
{\pi}^h(\uc) \REV{\doteq} \sum_a \sigma^a(\uc) \nedsh^a
\end{align}
where $\sigma^a(\uc)$ are the evaluation of the moments, i.e., the \ac{dof} values, and $\nedsh^a$ are the elements of the unique basis of functions that satisfies $\sigma_a(\nedsh^b) = \delta_{ab}$, i.e., the shape functions. The reader is referred to \cite{molm_nedelec} for a comprehensive definition of edge moments and the construction of polynomial spaces and basis of shape functions for the tetrahedral/hexahedral edge \ac{fe} of arbitrary order. 

\subsection{Objects}\label{subsec-globs}
In this section we introduce the definitions of global objects, or simply \emph{globs}, which are heavily used in DD preconditioners (see, e.g., \cite{DD_toselli}). Given a geometrical entity $s \subset \Gamma(\partition)$ and a subdomain partition $\partition$, we denote by $\neigho(s)$ the set of subdomains in $\partition$ that contain $s$. Then, we define a geometrical object as the maximal set $\lambda$ of geometrical entities in $\Gamma(\partition)$ with the same $\neigho(s)$ subdomain set. We denote by $\neigho(\lambda)$ the set of subdomains in $\partition$ containing $\lambda$ and by ndof($\lambda$) the total number of \acp{dof} placed \REV{on} $s\in\lambda$. An object $\lambda$ such that ndof($\lambda$)>0 is a face $F$ if $|\neigho(\lambda)|=2$ or an edge $E$ if $|\neigho(\lambda)|>2$. In addition, an object such that ndof($\lambda$)=0 is a corner. Grouping together the objects of the same type, we obtain the set of corners $\Lambda_C$, edges $\Lambda_E$ and the set of faces $\Lambda_F$. Therefore, the set of \emph{globs} is defined as $\Lambda(\partition) = \Lambda_C \cup \Lambda_E \cup \Lambda_F$.  

\begin{remark}
This definition differs from the standard one (see, e.g., \cite{badia_implementation_2013}).
It is intentionally done in order to isolate \emph{globs} that do not contain \acp{dof}, i.e., $\Lambda_C$, which can be omitted in the rest of our exposition. 
\end{remark} 

Once \emph{globs} are defined, let us also introduce the set of \ac{pb}-\emph{globs}, denoted by $\Lambda_{\rm pb}(\partition)$, as classification of all $s\subset\Gamma(\partition)$ into $\Lambda_C$, $\Lambda_E$ or $\Lambda_F$ by considering the previous definitions based on $\neigh(s)$ rather than $\neigho(s)$. $\Lambda_{\rm pb}(\partition)$ is a sub-partition of $\Lambda(\partition)$ where coefficients are subdomain-wise constant within each $\lambda \in \Lambda_{\rm pb}(\partition)$.   


\section{Physics-Based BDDC}\label{sec-pbbddc}
\subsection{Change of basis}\label{subsec-cb}

Any \ac{bddc} method that employs a standard 3D edge \ac{fe} basis of shape functions is bound to show a factor dependent on the element size $h^{-2}$ in the condition number \cite{toselli_dual-primal_2006}, which precludes scalability. 
A key aspect of the curl-conforming edge \ac{fe} spaces is the fact that $\grad \Vi \subset \Xi$. One of the main ingredients of any BDDC method are the averaging operators $\weiope: \X \rightarrow
\Xg$ (see detailed exposition in \sect{subsec-prec}) that restore the continuity of the solution at the interface among subdomains. Since the averaging operators are usually based on some algebraic operations over \ac{dof} values, they are, more precisely, scaling matrices that depend on the basis being used to describe $\X$ (and $\Xi$, by restriction to every subdomain). A key property that must hold \REV{for} such operator to end up with a 
stable decomposition is the following: Given a function $\uc \in \X$ such that its
local component in every processor belongs to $\grad \Vi$, the restriction of
the resulting function $\weiope \uc \in \Xg$ to every subdomain must belong to $\grad \Vi$ too. Otherwise, the energy of such functions is much increased after the averaging operation, and thus, the decomposition is not \emph{scalable}. A key result in this direction is the decomposition proposed in \cite{toselli_dual-primal_2006} in the \REV{framework} of FETI-DP methods for problems in $\Hcurl$.

Edge \ac{fe} space moments can be assigned to edges/faces of the mesh (see \cite{molm_nedelec}). Let us denote by $\Xi(I)$ the subspace of functions of $\Xi$ such that their \ac{dof} values are not \emph{located} on some $E \in \Lambda_E$ or $F \in \Lambda_F$, i.e., the \acp{dof} are interior. Clearly, $\Xi = \{\Xi(I)\} \oplus \{\Xi(F)\}_{F \subset \partial \tilde \Omega_i} \oplus \{ \Xi(E) \}_{E \subset \partial \tilde \Omega_i}$. On the other hand, a function $\vc^i \in \Xi(E)$ for a coarse edge $E \subset \partial
\tilde \Omega_i$ admits a unique decomposition as follows (see
\cite{toselli_dual-primal_2006,dohrmann_bddc_2016} for more details):
\begin{align} \label{eq:dec-edg}
  \vc^i \cdot \t_E = s_{0,E}(\vc^i) \Phi_E \cdot \t_E  + \sum_{j=1}^{n_E-1} w_{jE}(\vc^i) \grad \phi_E^j \cdot \t_E, \qquad \forall E \in \Lambda_E, 
\end{align} 
with $\lagsh_E^j \in \Vi (E)$ being the Lagrangian shape functions related to the internal nodes of $E$ (i.e., nodes $\xi\in E$ such that $\xi \notin \partial E$) and $n_E$ their cardinality, whereas $s_{0,E}(\vc^i) \doteq \int_E \vc^i \cdot \t_E ds$. It is clear from \eq{eq:dec-edg} that two kinds of \acp{dof} arise in the new basis for each subdomain edge $E\in \Lambda_E$: a \ac{dof} associated with the basis function $\Phi_E$, which represents the average tangent value over the coarse edge $E$, and \acp{dof} associated with gradients of scalar, Lagrangian shape functions placed at its internal nodes. An illustration for the variables in the old (original) and new basis for a given $E$ is presented in \fig{fig-bases}. Thus, we have that $\Xi(E) \doteq \grad \Vi(E) \oplus \Phi_E $. As a result, $\Xi$ admits the unique decomposition: 
\begin{align}\label{eq:spa-dec}
  \Xi = \{\Xi(I)\} \oplus \{\Xi(F)\}_{F \subset \partial \tilde \Omega_i} \oplus \{ \grad
    \Vi(E) \}_{E \subset \partial \tilde \Omega_i} \oplus \{ \Phi_E \}_{E
  \subset \partial \tilde \Omega_i},
\end{align}
where $\Phi_E$ is the tangential vector such that $\Phi_E \cdot \t_E = 1$, \REV{$\t_E$ being} the unit tangent to $E \in \Lambda_E$. 

Let us now describe the relation between the original set of \acp{dof} (old basis in the global space) and the one that arises from \eq{eq:spa-dec} (new basis) for $\Xg$. A function $\uc \in \Xg$ can be written in the old basis as $ \uc = \sum_a u^a \nedsh^a$, where $\nedsh = \{\nedsh^1, \ldots, \nedsh^n\}$ is the set of global edge shape functions. Furthermore, consider the set of new basis functions $\boldsymbol{\psi} = \{\boldsymbol\psi^1, \ldots, \boldsymbol\psi^n\}$, where old basis elements $\nedsh^a$ associated to $E \in \Lambda_E$ are replaced by its corresponding functions in \eq{eq:dec-edg} (i.e., interior and face edge functions, Lagrangian shape functions gradients, and the coarse edge functions). 
The interpolation operator $\pi^h$ (see \eq{eq-int_op}) induces the change of basis matrix, whose entries are computed by evaluating the original edge moments $\sigma^a$ for the introduced set of new basis functions $\boldsymbol{\psi}$ as (Einstein notation) 
\begin{align}\label{eq-cb}
u_{\rm old}^a = \sigma^a (\uc_{\rm new}) = \sigma^a (\boldsymbol \psi^b) u^b_{\rm new} = \Cb_{ab} u^b_{\rm new}, 
\end{align} 
or in compact form, $\uc_{\rm old} = \Cb \uc_{\rm new}$. Furthermore, we can readily define the inverse change of basis as $\uc_{\rm new} = \Cbinv \uc_{\rm old}$. The usual restriction matrix $R_i: \Xg \rightarrow \Xi$ is used to obtain local restrictions of the global change of basis as $\Cbi = R_i \Cb R^T_i$; we abuse notation, using the same notation for the restriction with respect to the two bases, since it will be clear from the context. Finally, local restrictions lead to the change of basis ${\Cbw} = \prod_i \Cbi$, which will be applied for functions  defined on $\X$. 
A detailed exposition of an implementation strategy for the change of basis is found in \sect{subsec-cb_implementation}. 

\subsection{Preconditioner}\label{subsec-prec}
Similarly to other \ac{bddc} methods, we associate coarse \acp{dof} to some of the \emph{globs} in $\Lambda_{\rm pb}(\partition)$. In particular, \ac{bddc} methods for 3D curl-conforming spaces associate two coarse \acp{dof} to every $E \in \Lambda_E$, defined as  
 
\begin{subequations} 
\begin{align} 
s_{0,E}(\vc^i) & \doteq \int_E \vc^i \cdot \t_E ds,  \label{eq-cd1} \\   
s_{1,E}(\vc^i) & \doteq \int_E s \vc^i \cdot \t_E ds, \label{eq-cd2} 
\end{align} 
\end{subequations} 
where $s$ is an arc-length parameter, $s \in [ -|E|/2, |E|/2 ]$. Thus, the expression \eq{eq-cd2} refers to the first order moment of the tangent component of the solution on the edge $E$, in contrast to the zero-order moment in \eq{eq-cd1}. In the new basis (see \eq{eq:dec-edg}), it is easy to check that $s_{0,E}(\vc^i) =
s_{0,E}(\Phi_E)$ and $s_{1,E}(\vc^i) =
s_{1,E}(\sum_{j=1}^{n_E-1} w_{jE} \grad \lagsh_E^j)$ \cite{toselli_dual-primal_2006}.
 Let us define the subspace $\Xbddc$ as  
\begin{align}
\Xbddc \doteq \{ \w \in \X: s^\mathcal{D}_{0,E} = s^\mathcal{D'}_{0,E}, \, s^\mathcal{D}_{1,E}=s^\mathcal{D'}_{1,E} \ \forall E \in \Lambda_E, \forall \mathcal{D}, \mathcal{D}' \in \neigh(E) \} 
\end{align} i.e., the subspace $\Xbddc \subset \X$ such that for all $\w \in \X$, coarse \acp{dof} \eqref{eq-cd1} and \eqref{eq-cd2} are continuous across subdomain interfaces $\Gamma(\partition)$ for all $E \in \Lambda_E$. Clearly, $\Xg \subset \Xbddc \subset \X$.
 
The following key ingredient in the \ac{bddc} method is the averaging operator $\weiope: \X
\rightarrow \Xg$, defined as some weighted average of the \ac{dof} values at the interface. This operator is in practice defined as a matrix for a
particular choice of the basis functions for $\Xi$. Let us consider the new basis functions in $\boldsymbol{\psi}$. Given a fine edge/face $f \subset \Gamma(\partition)$, we define a weight for each $\mathcal{D} \in \neigho(f)$ as 
\begin{align}\label{eq-wop}
\delta_\mathcal{D}^\dagger(f) = \frac{\sum_{ \mathcal{D'} \in \neigh(f) \cap \mathcal{D}} \chi_{_\mathcal{D'}}}{\sum_{ \mathcal{D'} \in \neigh(f) } {\chi_{_\mathcal{D'}}} }, 
\end{align}
where the choice of $\chi$ defines the scaling: the \emph{cardinality} scaling with $\chi=1.0$ or the $\curlc$-based, $\massc$-based scaling with $\chi=\curlc$ or $\chi=\massc$, respectively. Besides, one can consider a weighted coefficient for $\chi$,  $\omega = \curlc + \massc h^2$, which is also subdomain-wise constant within all $E \in \Lambda_E$ in our definitions if regular structured meshes are considered. 
 We note that all the expressions for the scalings are constant on \emph{globs} by construction, due to objects generation based on $\pbpartition$ with constant coefficients. Then, we define the weighted function $\weiope \vc \in \Xg$ as follows. First, we compute for every subdomain \REV{$\Omega_i$} the weighted local functions as
\begin{align}
\w^i = \vc^i_I + \sum_{{F \subset \Lambda_F}} \delta_\mathcal{D}^\dagger(F) \vc^i_F + \sum_{E
\subset \Lambda_E}  \delta_\mathcal{D}^\dagger(E) \vc^i_E,
\end{align}
where $\vc^I$, $\vc^i_F$, and $\vc^i_E$ include the components related to interior, face, and edge \acp{dof} in \eq{eq:spa-dec}, respectively, and \REV{$\delta_\mathcal{D}^\dagger$ the corresponding weight for every interface edge/face $f$ in $\mathcal{D} \in \neigho(f) \cap \Omega_i$.}
Next, we sum the values of \acp{dof} on different subdomains that represent the
same \acp{dof} in $\X$, i.e., assemble the \acp{dof} as 
\begin{align} 
\vc = \sum_i R_i^T \w^i.
\end{align} 
Next, we recover  the sub-assembled bilinear form $\Ad$, whereas $\Ag$ and
$\Abddc$ are the Galerkin projection of $\Ad$ onto $\Xg$ and $\Xbddc$,
respectively. We additionally define the harmonic extension operator $\mathcal{E}$, that,
given $\uc \in \Xg$, provides $\uc + \delta \uc_I$, where $\delta \uc_I \in \Xi (I)
$ is a
bubble function that vanishes on the interface $\Gamma(\partition)$ and holds:
\begin{align}
\langle \Ai \delta \uc_I^i, \vc_I^i \rangle = - 
\langle \Ai \uc^i, \vc_I^i \rangle, \quad \forall \vc_I^i \in \Xi(I). 
\end{align}
Let us denote the Galerkin projection of $\Ag$ onto the global bubble space $\Xg(I)\doteq \{ \vc \in \Xg \text{ such that } \vc = \boldsymbol{0} \text{ on } \Gamma(\partition) \}$ by $\Ab$. Thus, the action of the harmonic extension operator can be written as 
\begin{align} 
\mathcal{E} \uc \doteq ( 1 - \Abinv \Ag ) \uc. 
\end{align} 
We finally define the operator $\mathcal{H} = \mathcal{E} \weiope$. We can now state the \ac{bddc} preconditioner as  
\begin{align}\label{eq-prec_new}
\preope \doteq \Abinv + \mathcal{H} (\Abddc)^{-1}
\mathcal{H}^T.
\end{align}
Note that having the expression of the operators associated with the new basis is essential in order to apply the averaging operator $\weiope$. Nevertheless, it is possible to employ the original operators in the standard basis and work with the change of basis matrix $\Cb$ \cite{dohrmann_recent_tools}. In this case, the only difference with regard to \eq{eq-prec_new} is the application of the averaging operator as 
\begin{align}\label{eq-cbapp}
\mathcal{H} = \mathcal{E} \Cb \weiope \Cbwinv \qquad \text{or} \qquad \mathcal{H}^T = \Cbwtinv \weiope^T \Cb^T \mathcal{E}^T. 
\end{align}  
Application details for the change of basis are exposed in \sect{subsec-cb_implementation}. Therefore, the definition of the preconditioner is the one of the standard BDDC \cite{dohrmann_preconditioner_2003} with a set of \emph{globs} generated by a partition based on coefficients and a modification of the averaging operator to take into account this fact. Besides, one can work with the standard basis of edge \acp{fe} and use strategically the change of basis required to attain a scalable algorithm in the application of the weighting operator. 
 
\subsection{Perturbed \ac{pbbddc} preconditioner}\label{subsec-pertur}
The presented \ac{pbbddc} preconditioner has been shown to be robust with the jump of coefficients in the steady Poisson equation \cite{BadiaNguyen_physics_based}. However, the problem in \eq{eq-strong_maxwell} adds the complexity of the interplay between the two different parameters $\curlc$ and $\massc$ across the interface. Following the robust approach in \cite{BadiaNguyen_physics_based}, our idea is to get rid of the jump of one coefficient across the interface so the preconditioner has not to deal with the interplay between the two of them and the scenario where the method is successful is recovered. In order to decide which coefficient is affected, we consider the locality of the mass matrix operator in front of the double curl terms. The main idea is to add a perturbation in the original formulation of the preconditioner so we end up with common information for the mass matrix operator for \acp{dof} that are replicated among different subdomains, i.e., located \REV{on} the interface $\Gamma(\partition)$. Therefore, the problem posed in $\Xbddc$ will only contain a jump in the double curl term across the interface. 

Given a function $\uc^i \in \Xi$, we can define its extension as a global function $\bar \uc^i \in \Xg$ such that all \acp{dof} belonging to $\tilde \Omega_i$ are identical to the ones of $\uc^i$ and the rest are zero. The extended function has support on $\tilde \Omega_i$ and its neighbours, denoted by $\bar{\Omega}_i$. The perturbed preconditioner for a local subdomain $\tilde \Omega_i$ is expressed as: 
\begin{align}\label{eq-massint}
\tilde \Ai( \uc^i, \vc^i) = \int_{\tilde \Omega_i}  (\curlc \curl \uc^i)\cdot(\curl \vc^i) dx + \int_{\bar \Omega_i} \massc \bar \uc^i \cdot \bar \vc^i  dx.
\end{align} 

Therefore, entries for interface \acp{dof} in the local mass matrix will be \emph{fully-assembled} instead of \emph{partially} assembled, leading to common information at the interface across all subdomains. In the situation where no jump occurs for the mass matrix coefficients at the interface among subdomains, we consider the original preconditioner presented in \sect{subsec-prec}, avoiding the perturbed formulation for obvious reasons. 

\begin{remark} 
The definition of the original problem is not modified. We only consider the perturbed local operator $\tilde \Ai$ in the formulation for the preconditioner. 
\end{remark}

\subsection{Relaxed \ac{pbbddc}} 
In previous sections, the definition of $\pbpartition$ (and consequently the definition of \ac{pb}-\emph{globs}) is based on the requirement that coefficients are constant in each \ac{pb}-subdomain, i.e., different subparts with constant coefficients can be identified in a subdomain, e.g. a problem composed by different homogeneous materials. However, physical coefficients may vary across a wide spectrum of values, even in a small spatial scale. Besides, the requirement that coefficients have to be constant in each \ac{pb}-subdomain may result in an over-partitioned domain where coefficient jumps are not significant among different \ac{pb}-subdomains. In order to address these situations and to deal with a more general applicability of the preconditioner, we introduce the \ac{rpbbddc} extension of the preconditioner. In short, \ac{rpb}-subdomains are not determined by constant coefficients within the original partition but we only require that the maximal contrast in each \ac{pb}-subdomain is less than some predefined tolerance $r$. We define the maximal contrast independently for each coefficient present in the problem \eq{eq-strong_maxwell}, thus defining two (different) thresholds. Then, one can find a \ac{rpb}-partition, which we denote by $\rpbpartition$, such that 
\begin{align}\label{eq-thresholds}
\frac{ \curlc_{\rm max}(\mathcal{D})}{ \curlc_{\rm min}(\mathcal{D})} < r_\curlc  \quad \text{and} \quad  \frac{ \massc_{\rm max}(\mathcal{D})}{ \massc_{\rm min}(\mathcal{D})} < r_\massc \qquad \forall \mathcal{D} \in \rpbpartition
\end{align}  
where $\{ r_{\curlc}, r_\massc \} \geq 1$. Hence, the choice of both thresholds will determine the partition $\rpbpartition$ as a sub-partition of the original partition $\partition$. Note that if we consider $r_\curlc = r_\massc = \infty$, we recover the original partition $\partition$, while lower values for the thresholds lead to an increasing number of subparts, consequently \emph{globs}, and thus richer coarse spaces. The \ac{rpbbddc} preconditioner can be defined for any value of the threshold $r > 1$. By tuning $r$ one can obtain the right balance between computational time and robustness.

As coefficients $\curlc, \massc$ are no longer constant in each \ac{rpb}-subdomain, we propose to use averaged coefficients in \eq{eq-wop} in order to define the averaging operator. The averaged coefficients, denoted as $\bar \curlc$ and $\bar \massc$, are computed in the \ac{rpb}-subdomain $\Omega_k$ simply as 
\begin{align}
\bar \curlc = \frac{1}{| \Omega_k | } \int_{ \Omega_k} \curlc dx,  & &  \bar \massc = \frac{1}{| \Omega_k | } \int_{ \Omega_k} \massc dx.
\end{align} 
Hence, the definition of the averaging operator \eq{eq-wop} is not modified and all \acp{dof} \REV{on} the same coarse geometrical entity are weighted by the same constant value. Thus, under the $\rpbpartition$, the preconditioner expression is written exactly in the same form as in \sect{subsec-prec}.  

\section{Implementation aspects}\label{sec-implementation}
 In this section, we expose implementation strategies for some key points that the authors find of interest for potential users/developers of similar methods, namely an edge partition algorithm to avoid problematic cases in (unstructured) \ac{pb}-partitions, the construction of the change of basis, the implementation of the original \ac{bddc} constraints and the aggregation of cells into \ac{rpb}-subdomains based on heterogeneous coefficients $\curlc$, $\massc$ and the thresholds $r_\curlc$, $r_\massc$.  
 For a comprehensive implementation strategy of arbitrary order curl-conforming tetrahedral/hexahedral \acp{fe}, the reader is referred to \cite{molm_nedelec}. 

\subsection{Coarse edge partition}\label{sec-E_part}
Special care has to be taken with the general definition for subdomain edges presented in \sect{subsec-globs}. In particular, when \emph{globs} are generated based on $\pbpartition$ or a partition obtained with graph partitioners, e.g. METIS, the presented definition of $E$ in \sect{subsec-globs} may not be sufficient for expressing the function and the coarse \acp{dof} in the new basis. We detail the pathological cases identified in \cite{dohrmann_bddc_2016} (cases [1] and [2] below), and extension of a case in  \cite{dohrmann_bddc_2016} (case [3]) plus an additional case (case [4]), for which we provide examples. We propose a unique cure, based on the partition of problematic coarse edges $E$ into coarse sub-edges $E_j$ such that the problematic cases are solved. 

\begin{enumerate}[label={[\arabic*]}]\label{enum-path_cases}
\item \emph{Disconnected components.}
We say that fine edges $e \in E$ are connected if they have an endpoint in common. Consequently, if a coarse edge $E$ has $m$ disconnected components, it has $2m$ endpoints. Note that, while this fact does not preclude the invertibility of the change of basis, if each of the components is treated as a coarse edge we recover original meaningful definitions for continuity constraints across subdomains. 

\item \emph{Interior node in touch with another subdomain.} This case occurs when an internal node $v$ to $E$ does not have the same set of subdomains as $\neigh(E)$, i.e., is shared by $\neigh(E)$ plus additional subdomains. In fact, $v$ is then an element of $\Lambda_C$ in the classification provided in \sect{subsec-globs}. We recall that the change of variables is made for gradients of scalar, Lagrangian functions $\grad \lagsh_E$ defined on all internal nodes of $E$. However, if we consider a nodal shape function associated to $v$, it will be coupled with other internal nodal \acp{dof} for $E$, thus introducing a coupling between an external subdomain to $\neigh(E)$ and itself, which is clearly not present in the original basis. A remedy for it consists \REV{of} simply splitting the coarse edge $E$ into two sub-edges at the problematic node $v$. Let us denote by $V_p$ the subset of this kind of nodes for all $e \in E$. 

\item \emph{Edge $n$-furcation.}
This situation occurs when a coarse edge $E$ that does not have disconnected components has more than two endpoints. At some internal node the coarse edge is $n$-furcated into $n$ edges, so the definition of the shape function $\Phi_E$ in the new basis loses its original meaning. Furthermore, this fact precludes the locality of the change of basis for every edge $E$. In this case, a simple remedy is again to split the edge into sub-edges at any node shared by more than two edges.

\item \emph{Closed loop.} In this case we cannot identify endpoints for a coarse edge and therefore define a unique orientation for it. Furthermore, the new set of basis functions is not well defined since the definition in \sect{subsec-cb} relies on the fact that every edge has 2 end points, thus not being applicable in this case. In this situation, an internal node for the coarse edge $E$ must be chosen as start/end point (common in all subdomains) to assign an orientation to the edge and be treated as an edge endpoint in the change of basis definition. 
\end{enumerate} 

In order to address all the presented problematic cases we propose a simple algorithm based on a classification $\forall e \in E$ into sub-edges. Our goal is to find a partition of $e \in E \in \Lambda_E$ into $E_j$ such that every $E_j$ is constructed connecting fine edges that share (only) one vertex with the following edge. Therefore, every coarse sub-edge $E_j \subset E$ has a unique starting point, a chain of connected \emph{fine} edges sharing only one node and a unique end-point, which defines its unique orientation across all subdomains. Let us consider the set of nodes $V = \cup_{e \in E } (v \in \partial e)$, where the number of occurrences for each node $v\in V$ is denoted by count($v$). First, we can identify the set of nodes where $E$ is $n$-furcated as 
\begin{align} 
\text{$n$-furcation nodes}      & \quad V_N  \doteq \{ v \in V\setminus V_p | \quad {\rm count}(v)>2 \} \label{eq-s3} 
\end{align}
We note that $V_p$ is already identified in the \emph{glob} generation algorithm. Then, we can find a partition of the set of nodes into the two following subsets: 
\begin{subequations}\label{eq-subsets}
\begin{align} 
\text{Edge boundary nodes} & \quad V_B  \doteq \{ v \in V | \quad {\rm count}(v)=1 \} \cup {V_p} \cup {V_N} \label{eq-s1}\\ 
\text{Interior nodes}      & \quad V_I  \doteq \{ v \in V\setminus V_B \}. \label{eq-s2}
\end{align} 
\end{subequations}   
Note that by definition of interior nodes, they are such that ${\rm count}(v)=2$. Such classification is performed by simply counting the number of appearances of nodes plus setting problematic nodes belonging to other objects as edge boundary nodes. Then, the coarse edge partitioning Alg. \ref{alg-edge_part} finds paths from one edge boundary node (with a global criteria to select it) \REV{until} the following edge boundary node. Furthermore, in this procedure we identify the direction of every fine edge with regard to its container coarse edge. 

\begin{algorithm}
\small{ 
\KwData{$V = V_B \sqcup V_I$, $e \in E$}\KwResult{$E_j$ s.t. $E = \sqcup E_j$ }
 $ j \leftarrow 0$ \\
\While{ {\rm card}($V$) > 0 }{
        \uIf{ {\rm card}($V_B$) > 0}{
          Find $v^s \in V_B$ with minimum global id } 
        \Else{
          Find $v^s \in V_I$ with minimum global id \\
          $V_I \leftarrow V_I \setminus v^s$ and $V_B \leftarrow V_B \cup v^s$}
  $ j \leftarrow j+ 1$ \\
  Find $v^e$ s.t. $\{ v^s, v^e \} \in \partial e$ with minimum global id \hspace{0.5cm}   $ E_j \leftarrow \{ e \}$\\
  \textit{update counters and subsets} (Alg. \ref{alg-update_counter}) \\

  \While{ $v^e \in V_I$}{
  $v^s \leftarrow v^e$ \\
  Find $v^e$ s.t. $\{ v^s, v^e \} \in \partial e$ \hspace{0.5cm} $E_j \leftarrow E_j \cup e$ \\ 
   \textit{update counters and subsets} (Alg. \ref{alg-update_counter}) }}

}
\caption{Edge partition algorithm}
\label{alg-edge_part} 
\end{algorithm}


\begin{algorithm}
\small{ 
\KwData{$v^s, v^e, V_B, V_I$} \KwResult{$V_B, V_I$ }

\For{ $k$ in $s,e$ }{
count($v^k) \leftarrow {\rm count}(v^k) - 1$ \\ 
        \uIf{ count($v^k$) = 0 .and. $v^k \in V_B$}{
          $V_B \leftarrow V_B \setminus v^k$  } 
        \ElseIf{ count($v^k$) = 0 .and. $v^k \in V_I$ }{
          $V_I \leftarrow V_I \setminus v^k$}
}
}
\caption{Update counters and subsets}
\label{alg-update_counter} 
\end{algorithm}

From this point onwards, we consider that each $E \in \Lambda_E$ is a (sub-)edge of the original coarse edges such that they do not present problematic cases. 

\subsection{Change of basis}\label{subsec-cb_implementation} 

In this section we provide some implementation details of the change of basis described in \sect{subsec-cb}. In the application of the averaging operator in \eq{eq-cbapp}, we note that one must apply the global change of basis and its restriction to subdomains. A practical implementation of their application in both cases can be performed with the local restriction of the change of basis to the subdomains, i.e., $\Cbi = R_i \Cb R_i^T$, thus it can be performed in parallel in distributed memory environments. 
\REV{The application of the inverse of the change of basis in the sub-assembled space $\Cbwinv$ or $\Cbwtinv$ (\eq{eq-cbapp}) can be performed in parallel, i.e., relying on the restriction of the operators to the subdomains, given its definition (see \sect{subsec-cb}).}
On the other hand, the sparsity pattern of the global change of basis $\Cb$ can be exploited in order to achieve a parallel implementation of the application of $\Cb$ and $\Cb^T$ to a function $\uc \in \Xg$ that only relies on restricted (to the subdomains) information.
  
\begin{figure}[t!]
    \centering
    \begin{subfigure}[t]{0.23\textwidth}
        \includegraphics[width=\textwidth]{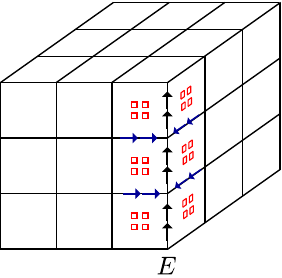}
        \caption{Black arrows represent standard basis \acp{dof} over $E$. }
        \label{fig-old_basis}
    \end{subfigure} \hspace{1cm}
    \begin{subfigure}[t]{0.23\textwidth}
        \includegraphics[width=\textwidth]{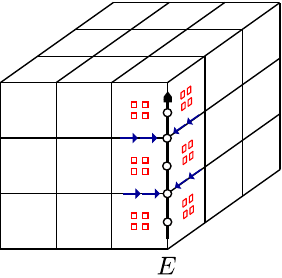}
        \caption{Nodes represent Lagrangian-like new basis \acp{dof}. Long arrow represents the \ac{dof} for function $\Phi_E$. }
        \label{fig-new_basis}
    \end{subfigure} 
    \caption{Standard (old) and new basis \acp{dof} for 3D hexahedra second order edge \ac{fe} over $E$. Additional depicted \acp{dof} values are affected by the change of basis for $E$, while the remaining \acp{dof} are invariant under the change of basis. }\label{fig-bases}
\end{figure}

\begin{proposition}\label{prop-Qu}
The expression $R_i \Cb \uc = \Cbi \uc^i$ holds, where $\uc \in \Xg$ and $\uc^i \in \Xi$.
\end{proposition} 
\begin{proof}
By definition of the change of basis matrix, it is easy to check that $R_i \Cb \uc$ only depends on the \ac{dof} values of $\uc^i$ (in the new base), i.e., $R_i \Cb \uc = R_i \Cb R_i^T R_i \uc$.
Thus we can write 
\begin{align}
R_i \Cb \uc &= R_i \Cb R_i^T R_i \uc = \Cbi \uc^i. 
\end{align} 
\end{proof} 
Unfortunately, this reasoning cannot be applied to the transpose of the change of basis. 
\begin{proposition} 
Consider arbitrary local weighting diagonal matrices $W_j$ for every subdomain such that $\uc = \sum_j R_j^T W_j R_j \uc$, i.e., they form a partition of the unity. 
Then, the expression $R_i \Cb^T \uc = R_i \sum_j R_j^T \Cbj^T W_j \uc^j$ holds, where $\uc \in \Xg$ and $\uc^j \in \Xj$.
\end{proposition}
\begin{proof}  
  Using the fact that $\sum_j R_j^T W_j R_j$ is the identity matrix and the fact that $R_j \Cb = R_j \Cb R_j^T R_j$ (as above), it holds:
\begin{align}
R_i \Cb^T \uc &=  R_i \sum_j \Cb^T R_j^T W_j R_j \uc 
= R_i \sum_j R_j^T R_j \Cb^T R_j^T W_j R_j \uc = R_i \sum_j R_j^T \Cbj^T W_j \uc^j.
\end{align} 
\end{proof} 
Therefore, the application of the change of basis can rely only on restrictions of the same to subdomains whereas the application of the transpose change of basis can be performed in parallel with subdomain restrictions plus nearest neighbour communications. 
 With this purpose in mind, we detail here how to implement the restriction of the change of basis local to subdomains. Let us define a partition of the \acp{dof} in ${\Xi}'$ into three subsets of \acp{dof}, namely: the \acp{dof} placed \REV{on} $e \in E$, the \acp{dof} placed \REV{on} interface edges/faces $f \notin E$ such that $\partial f \cap E \neq \emptyset$, and the remaining \acp{dof} in ${\Xi}'$, denoted by $\uc_E$, $\uc_{f}$ and $\uc_r$, respectively. Furthermore, let us consider that \acp{dof} in $\uc_E$ are sorted such that \acp{dof} belonging to the same coarse edge $E$ are found in consecutive positions.
Note that shape functions associated to $\uc_f$, $\uc_r$ are common in both (i.e., old and new) bases. For edge \acp{fe} of order $k$, $k$ moments (i.e., \acp{dof}) are defined \REV{on}
each $e \in \tri$. Let us denote by $n_e$ the number of fine edges $e \in E$. Then, the total number of \acp{dof} \REV{on} a coarse edge $E$ is $k n_e$. On the other hand, the number of Lagrangian-like \acp{dof} interior to $E$ (i.e., excluding $\partial E$) is $(k n_e - 1)$, i.e., the number of shape functions of the type $\grad \lagsh_E$. The change of basis is completed with the addition of the function $\Phi_E$ to the new basis so that the dimension of both bases coincides. For the sake of illustration, both sets of basis functions restricted to $E$ are depicted in \fig{fig-bases}. 

Let us denote by $n_E$ the number of coarse edges $E \in \Lambda_E$ for a given subdomain. Then, we define the change of basis $\mathcal{Q}^{E_j}$, $j=\{1,\ldots,n_E\}$  local to every $E_j \in \Lambda_E$ as  
\begin{align}\label{eq-Qe}
  \mathcal{Q}^{E_j}_{ab} = \sigma_a (\grad \lagsh_E^b), \quad {\rm for }\quad b=1,\ldots,kn^{E_j}_e-1 \qquad   \mathcal{Q}^{E_j}_{a,k n_e^{E_j}} = \sigma_a ( \Phi_{E_j} ),   
\end{align} 
where $\sigma_a$, $a=1,\ldots, k n^{E_j}_e$, are the (original basis) edge moments defined \REV{on} $E_j$ (the superscript in $n_e^{E_j}$ has been introduced to show that it depends on the coarse edge). We can now define the change of basis $\mathcal{Q}^E = {\rm diag}(\mathcal{Q}^{E_1}, \ldots, \mathcal{Q}^{E_{n_E}} )$ local to coarse edges. 
 We remark that the same orientation for every coarse edge $E \in \Lambda_E$ must be defined on the set of subdomains $\mathcal{D} \in \neigho(E)$. Otherwise, the definition of the new basis function $\Phi_E$ is not consistent across subdomains. 
In addition, the change of basis must take into account the effect of the new \ac{dof} values $\uc_E$ associated to $\grad \lagsh_E^j$ and $\Phi_E$ for $E$ in the old values $\uc_f$. Thus, we evaluate for all indices $b$ of shape functions associated to $\uc_E$   
\begin{align}\label{eq-qfc}
\mathcal{Q}^f_{cb} = \sigma_c (\grad \lagsh_E^b), 
\end{align} 
where $c$ corresponds to the index of all moments associated to $\uc_f$. We recall that, by definition, $\sigma_c(\Phi_E)=0$. The application of the moments $\sigma_c$ to the (original) shape functions associated to $\uc_f$ results in $\sigma_c(\nedsh^b) = \delta_{cb}$. 
Finally, \acp{dof} in $\uc_r$ are invariant under the change of basis. \REV{Note that the definition of Eqs. \eqref{eq-Qe} and \eqref{eq-qfc} related to the gradients of the scalar shape functions coincides with the so-called discrete gradient operator related to these functions, as used in \cite{Zampini2017}. However, we prefer to motivate the change of basis with the usage of the N\'ed\'elec interpolator, since it naturally provides the definition of the entries related to the unit tangent function, while suitable eigenvectors to complete the change of basis are computed in \cite{Zampini2017}. The structure of the change of basis restricted to a subdomain is }

\begin{align}\label{eq-cbmat}
\uc_{\rm old} = 
\begin{bmatrix} 
\uc_E \\
\uc_f \\ 
\uc_r
\end{bmatrix}_{\rm old}
 = 
\begin{bmatrix} 
\mathcal{Q}^E & 0 & 0 \\
\mathcal{Q}^f & I & 0 \\ 
         0  & 0 & I 
\end{bmatrix}
\begin{bmatrix} 
\uc_E \\
\uc_f \\ 
\uc_r
\end{bmatrix}_{\rm new} = \Cbi \uc_{\rm new},
\end{align} 
where it becomes clear the fact that the inverse of the change of basis is well defined if and only if $\mathcal{Q}^E$ is invertible. In turn, $\mathcal{Q}^E$ will be invertible if and only if every change of basis local to $E \in \Lambda_E$ (\eq{eq-Qe}) is invertible. 

\begin{remark} 
Although it is used in this exposition for the sake of clarity, we do not require any particular ordering of \acp{dof} in a practical implementation of $\mathcal{Q}_i$. 
\end{remark}  
    
\subsection{\ac{bddc} constraints}

In this subsection we propose a practical manner of computing the \ac{bddc} constraints Eqs.~(\ref{eq-cd1}) and (\ref{eq-cd2}) for local problems. In our implementation, constraints over local problems are strongly imposed through the usage of Lagrange multipliers on the original basis. Therefore, the local matrix $\Ai$ is extended with the discrete version of the constraints $C$ in order to obtain constrained (Neumann) local problems.

The computation of constraints requires to integrate zero and first order moments for the solution over all coarse edges $E$. We note that the first constraint \eq{eq-cd1} can be easily implemented for $k$-order edge \acp{fe} as 
\begin{align}
s_{0,E}(\uc^i) & \doteq \int_E \uc^i \cdot \t_E ds =  \int_E (\sum_{a=1}^{k n_e} u^a \nedsh^a) \cdot \t_E
ds = \sum_{a=1}^{k n_e} u^a \int_E \nedsh^a \cdot \t_E ds \label{eq-zeroint} \\ & = \sum_{a=1}^{k n_e} u^a \int_E \nedsh^a \cdot \t_E (\sum_{b=1}^k p_b) ds =  \sum_{a=1}^{k n_e}  (\t_e \cdot \t_E ) u^a   = \sum_{a=1}^{k n_e}  C^{a} u^a, \nonumber  
\end{align} 
where we used the partition of the scalar, unit function into the set of Lagrangian test functions $p_b$ belonging to the polynomial space $\V(E)$ of order $k-1$. These functions are used for defining the $k$ local moments on every $e$ as $\sigma^b(\uc) = \int_e \uc \cdot \t_e p_b$, $b=\{1, \ldots, k\}$ (see \cite{molm_nedelec} for details). Its duality with basis shape functions, i.e., $\sigma^b(\nedsh^a) = \delta_{ba}$, has been used in \eq{eq-zeroint}. 
Thus, to compute the first constraint one only needs to add $\pm 1$ at the corresponding entry in $C$ for each \ac{dof}, where the sign is determined by the agreement between fine and coarse edge orientations, i.e., $\t_e \cdot \t_E$. On the other hand, the computation of the second constraint \eq{eq-cd2} requires to define an arc-length parameter over $E$. A practical implementation of the constraint \eq{eq-cd2} can avoid it by considering the constraint in the new basis. Since $\lagsh_{E}^j$ vanish at $\partial E$, integration by parts yields \cite{toselli_dual-primal_2006}
\begin{align} 
s_{1,E}(\uc^i) & \doteq \int_E s \uc^i \cdot \t_E ds = \int_E  s (u_E \Phi_E + \sum_{a=1}^{k n_e -1} u^a \grad \lagsh_E^a) \cdot \t_E ds= \\ 
& -\sum_{a=1}^{k n_e-1}  u^a \int_E \lagsh_E^a ds  = -\sum_{a=1}^{k n_e -1} C_{\rm new}^a u^a, \nonumber 
\end{align} 
where the contribution of $\Phi_E$ is null due to the antisymmetry of the product $s \Phi_E$ ( we recall that $s \in [-|E|/2, |E|/2]$ ) over $E$. 
Then, we can apply the change of basis to obtain the expression in the original basis, i.e., $ C = C_{\rm new} \Cbiinv $.  

\subsection{Building $\rpbpartition$}\label{subsec-rpb_imp}
In the \ac{rpbbddc} method, a $\Theta_{\rm pb}^r$ partition is used such that the maximal contrast, for each one of the coefficients, is lower than a predefined tolerance $r$ in each subdomain. Our goal is to identify a partition of every subdomain into $\mathcal{D}'\in \rpbpartition$ subdomains where the thresholds $\frac{\curlc_{\rm max} (\mathcal{D}')}{\curlc_{\rm min} (\mathcal{D}')}<r_{\curlc}$ and $\frac{\massc_{\rm max} (\mathcal{D}')}{\massc_{\rm min} (\mathcal{D}')}<r_{\massc}$ are respected. It can be accomplished using different algorithms. One approach is to consider a seed cell and aggregate the surrounding cells such that the contrast(s) are below the given threshold(s), proceeding recursively \REV{until} no neighbouring cells can be aggregated. We take another seed among the non-aggregated cells and proceed again \REV{until} all cells have been processed. 
 
Alternatively, one can first determine the maximum and minimum values for $\curlc$ and $\massc$ in a given subdomain $\mathcal{D} \in \partition$. With this information and the thresholds $r_\curlc$, $r_\massc$, we can determine the number of sub-intervals for every subdomain and coefficient as follows. First, we compute $\ell_\curlc(\mathcal{D})$ and $\ell_\massc(\mathcal{D})$ as the smallest positive integers for which 
\begin{align} 
 \frac{\curlc_{\rm max}(\mathcal{D})}{\curlc_{\rm min}(\mathcal{D})} < r_\curlc^{\ell_\curlc(\mathcal{D})},  &  &\frac{\massc_{\rm max}(\mathcal{D})}{\massc_{\rm min}(\mathcal{D})} < r_\massc^{\ell_\massc(\mathcal{D})},
\end{align}
respectively. One can now define the intervals 
\begin{align}\label{eq-interv}
I_{i,j} \doteq [ r_\curlc^{i-1} \curlc_{\rm min}(\mathcal{D}), r_\curlc^{i} \curlc_{\rm min}(\mathcal{D}) ] \times [ r_\massc^{j-1} \massc_{\rm min}(\mathcal{D}), r_\massc^{j} \massc_{\rm min}(\mathcal{D}) ],
\end{align}
for $i \in [1,\ell_\curlc(\mathcal{D})]$, $j \in [1,\ell_\massc(\mathcal{D})]$. Cells with their coefficients on the same interval $I_{i,j}$ are aggregated in a \ac{pb}-subdomain. Those cells that have coefficients across multiple sub-intervals are treated as additional \ac{pb}-subdomains. \REV{This definition allows one to isolate cells that contain abrupt jumps in the value of the coefficients, while the user has the freedom to select the thresholds such that unnecessary extra PB-subdomains are avoided. } 

\begin{figure}[ht!]
      \centering
    \begin{subfigure}[t]{0.30\textwidth}
      \includegraphics[width=\textwidth]{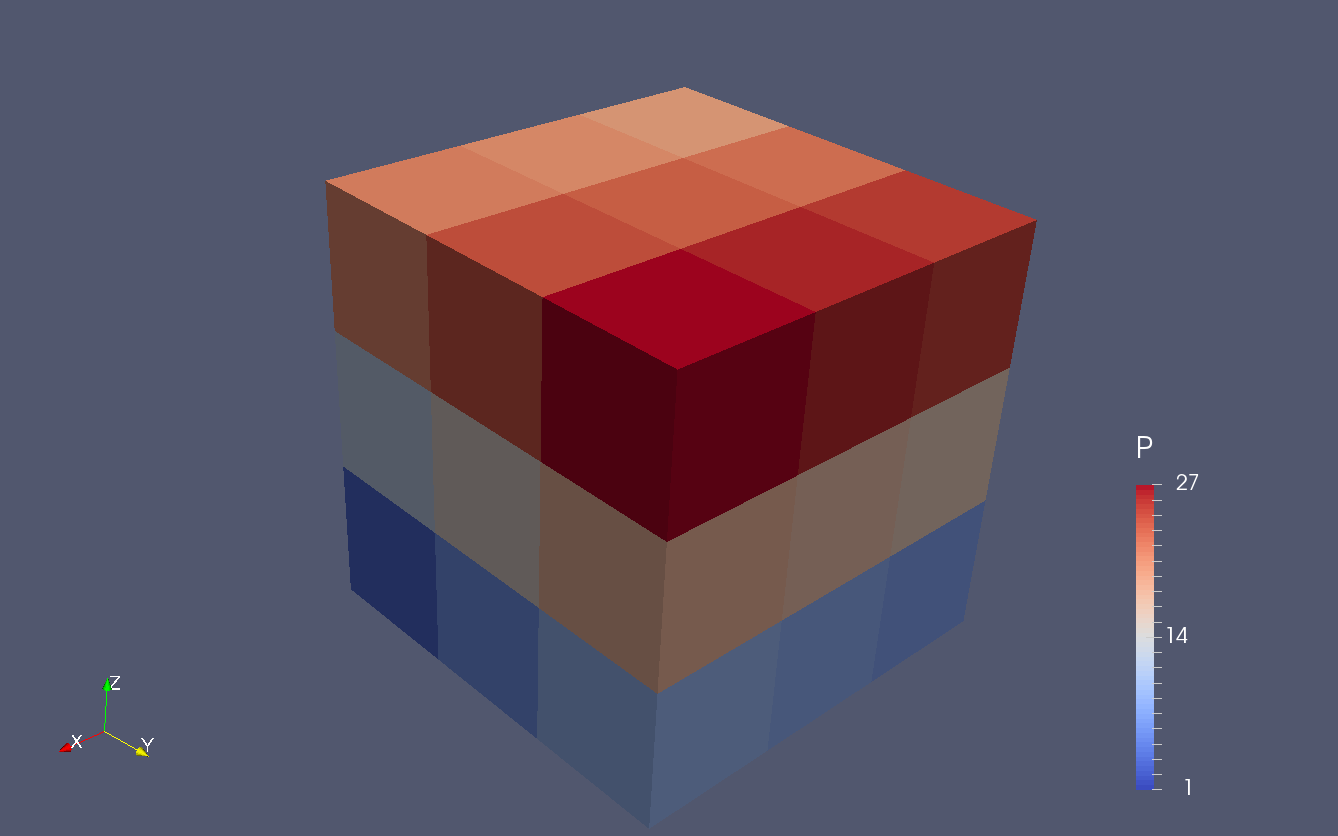}
      \caption{Original geometrical partition $\partition$}
      \label{fig-geom_partition} 
  \end{subfigure}
  \hspace{0.2cm}
  \begin{subfigure}[t]{0.30\textwidth}
    \includegraphics[width=\textwidth]{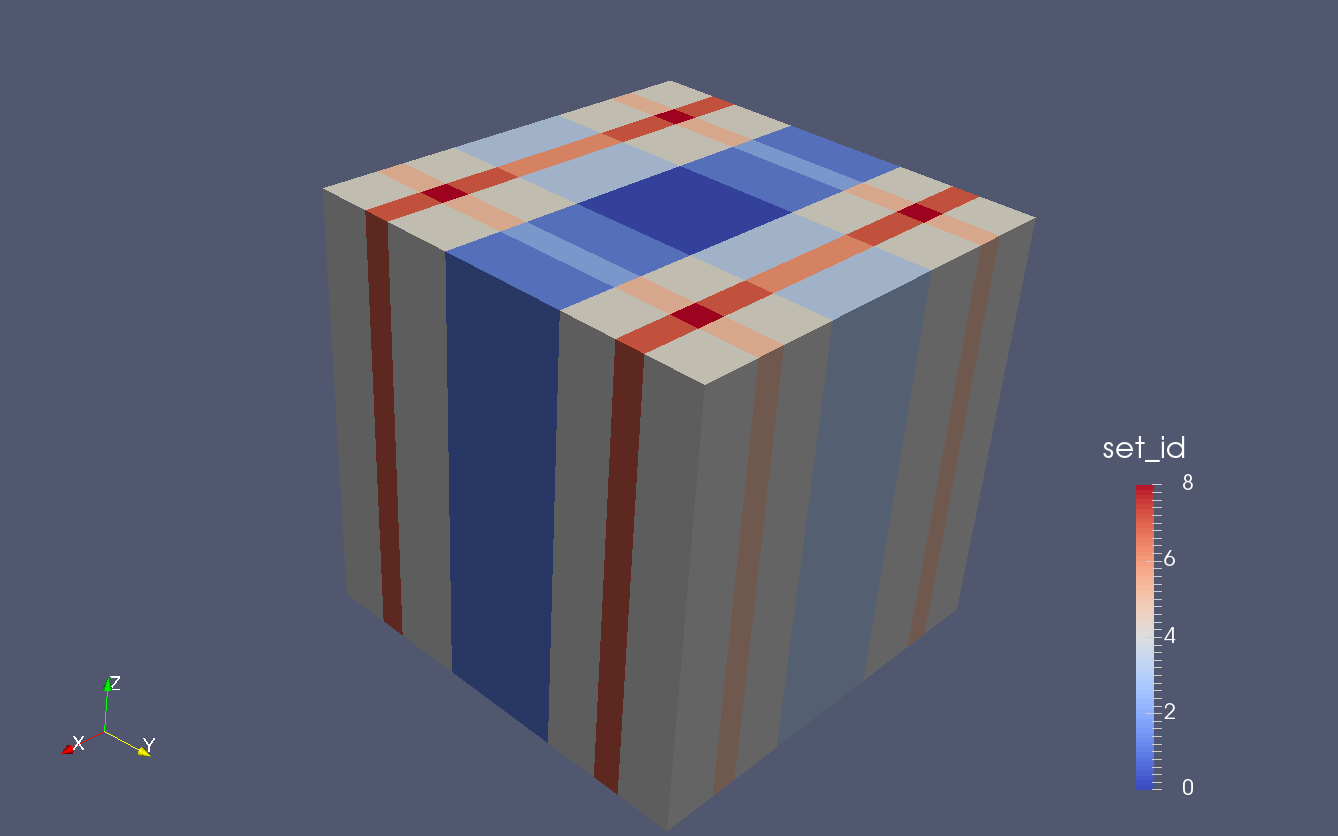}
    \caption{ Aggregation of cells into subsets based on coefficients }
    \label{fig-coeff_part}
  \end{subfigure} 
  \hspace{0.2cm}
    \begin{subfigure}[t]{0.30\textwidth}
    \includegraphics[width=\textwidth]{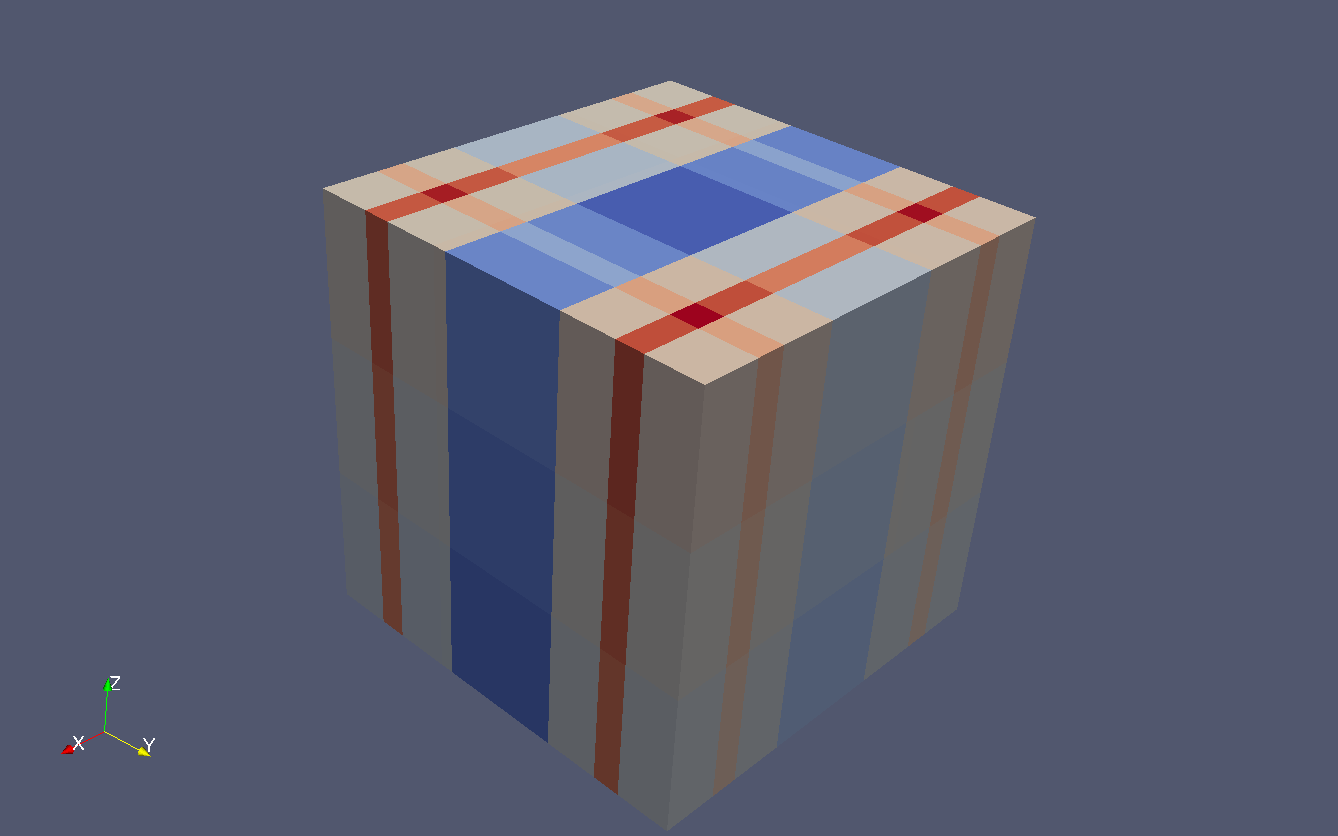}
    \caption{Resulting \ac{rpb}-partition $\Theta_{\rm pb}^r$.}
    \label{fig-pb_part}
  \end{subfigure} 

  \caption{Partitions for scalar coefficients described by $\log(\curlc)= 3 \sin( 3\pi x)$ and $\log(\massc)= 3 \sin( 3\pi y)$ with an initial $3\times 3\times 3$ partition of the unit cube.}
  \label{fig-partition_seqs}
\end{figure}

\begin{figure}[ht!]
       \centering
    \begin{subfigure}[t]{0.30\textwidth}
      \includegraphics[width=\textwidth]{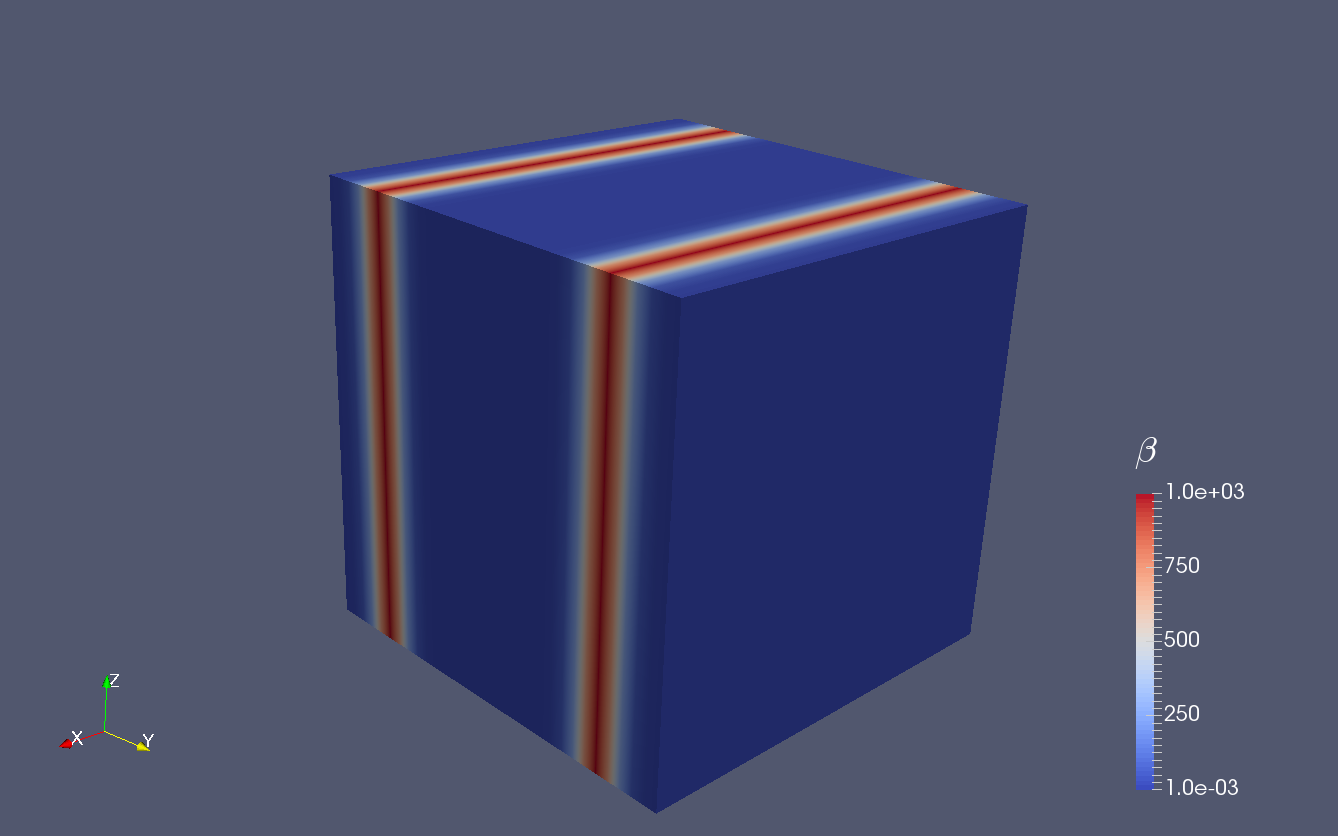}
      \caption{Analytical function.}
      \label{fig-beta_fun} 
  \end{subfigure}
  \hspace{0.2cm}
  \begin{subfigure}[t]{0.30\textwidth}
    \includegraphics[width=\textwidth]{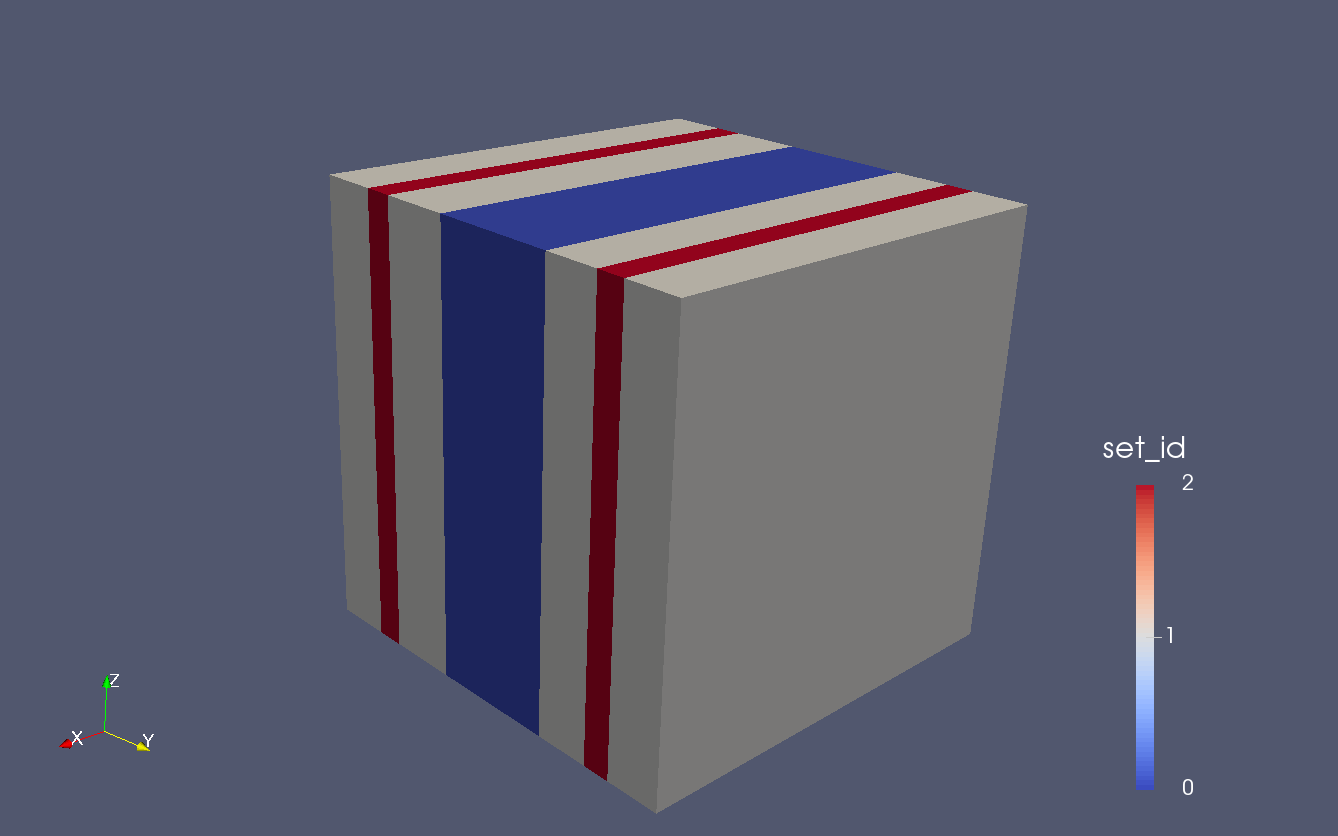}
    \caption{Aggregation of cells into subsets.}
    \label{fig-beta_sets} 
  \end{subfigure} 
  \caption{Aggregation of cells into subsets based on $log(\massc)= 3 \sin( 3\pi y)$ with threshold $r=10^3$.}
  \label{fig-heterogeneous_subsets}
\end{figure}

For the sake of illustration, we include an example where all the different subset indices are presented for a unit cube domain: the original  partition into $P=3\times 3\times 3$ subdomains in \fig{fig-geom_partition}, the aggregation of cells into subsets based on $\log(\massc) = 3\sin(3\pi y)$ (see \fig{fig-beta_fun}) for $r=10^3$ in \fig{fig-beta_sets}, combined with an analogous partition for $\log(\curlc)=3\sin(3\pi x)$ leading to a coefficient-based partition in \fig{fig-coeff_part} and the final \ac{rpb}-partition $\Theta_{\rm pb}^r$ in \fig{fig-pb_part}.   

\REV{We note that the preconditioner is very robust despite the presence of subdomains with large aspect ratios. Support to this observation can be found in \cite{BadiaNguyen_physics_based} (see Remark 3.12 and the corresponding numerical experiments).}


\section{Numerical results}\label{sec-results}

In this section we evaluate the weak scalability of the proposed preconditioner for the problem in \eq{eq-strong_maxwell}, within the preconditioned \ac{cg} Krylov iterative solver. The robustness of the \ac{rpbbddc}-CG solver is tested in 3D simple domains, which are discretized either with structured \REV{or} unstructured meshes. As performance metrics, we focus on the number of \ac{rpbbddc} preconditioned CG iterations required to attain the convergence criteria, which is defined as the reduction of the initial residual {algebraic} $\ell_2$-norm by a factor  $10^{-6}$. On the other hand, the total computation time will be presented, which will include both preconditioner set-up and the preconditioned iterative solution of the linear system in all the experiments reported. The particular definition of coefficients $\curlc$ and $\massc$ and its distribution will be specified throughout the section for each case. 

\REV{We have also applied a plain CG solver for the different tests, but the results of the latter are not reported here due to its poor performance. As an example, for the same problem set-up as in Fig.~\ref{fig-checks_weak}, only 48 processors, and $H/h=30$, it did not satisfy the convergence criteria within a prescribed maximum of 20,000 iterations, for which it consumed 144 secs, far beyond the proposed \ac{rpbbddc} preconditioner.}
 
\subsection{Experimental framework}

The \ac{rpbbddc} methods have been implemented in the scientific software project \FEMPAR~\cite{fempar-web-page}. \FEMPAR, developed by the Large Scale Scientific Computing (LSSC) team at CIMNE-UPC, is a parallel hybrid OpenMP/MPI, software package for the massively parallel FE simulation of multiphysics problems governed by PDEs. \FEMPAR offers a set of flexible data structures and algorithms for each step in the simulation pipeline, which can be customized in order to meet particular application problem needs. See \cite{badia-fempar} for a thorough coverage of the software architecture of \FEMPAR. Among other features, it provides the basic tools for the efficient parallel distributed-memory implementation of substructuring \ac{dd}  solvers~\cite{badia_implementation_2013,badia_multilevel_2016}, based on a \emph{fully-distributed} implementation of data structures involved in the parallel simulation. The parallel codes in \FEMPAR heavily use standard computational kernels provided by BLAS and LAPACK. Besides, through proper interfaces to several third party libraries, the local constrained Neumann problems and the global coarse-grid problem can be solved via sparse direct solvers. \FEMPAR is released under the GNU GPL v3 license, and is more than 300K lines of Fortran200X code long following object-oriented design principles. In this work, we use the overlapped BDDC implementation proposed in \cite{badia_scalable_2014}, with excellent scalability properties. It is based on the overlapped computation of \emph{coarse} and \emph{fine} duties. {As long as} coarse duties can be fully overlapped with fine duties, perfect weak scalability can be attained. We refer to \cite{badia_multilevel_2016} for more details.  

The experiments in this section have been performed
on the MareNostrum-IV \cite{MNIV} (MN-IV) supercomputer, hosted by the Barcelona Supercomputing Center (BSC). 
In all cases, we consider a one-to-one mapping among subdomains, cores and MPI tasks. Provided that the algorithm allows for a high degree of overlapping between fine and coarse duties, an additional MPI task is spawn into a full node (i.e., 48 cores) in order to perform the coarse problem related tasks. The multi-threaded PARDISO solver in Intel MKL is used to solve the coarse-grid problem within its computing node. 

 Unless otherwise stated, the problem \eq{eq-strong_maxwell} will be solved in the unit cubic domain $\Omega = [0,1]^3$ with Dirichlet homogeneous boundary conditions on the whole boundary and the forcing term $\f=1$. Let us denote by $h$ the usual mesh element size, and by $H$ the size of the subdomain. Then, local problem sizes can be characterized in a structured mesh and partition by $\frac{H}{h}$. In order to perform a weak scalability analysis, we build a set of structured meshes consisting on $(4 \frac{H}{h} k \times 4\frac{H}{h} k \times 3 \frac{H}{h} k) $ hexahedra. A uniform partition of the meshes into $ P=(4k\times 4k\times 3k)=48k^3$ subdomains is considered, where local problem sizes are  $(\frac{H}{h})^3$.  

\subsection{Homogeneous problem} 

\begin{figure}[ht!]
    \begin{subfigure}[b]{0.45\textwidth}
      \centering
      \includegraphics[width=\textwidth]{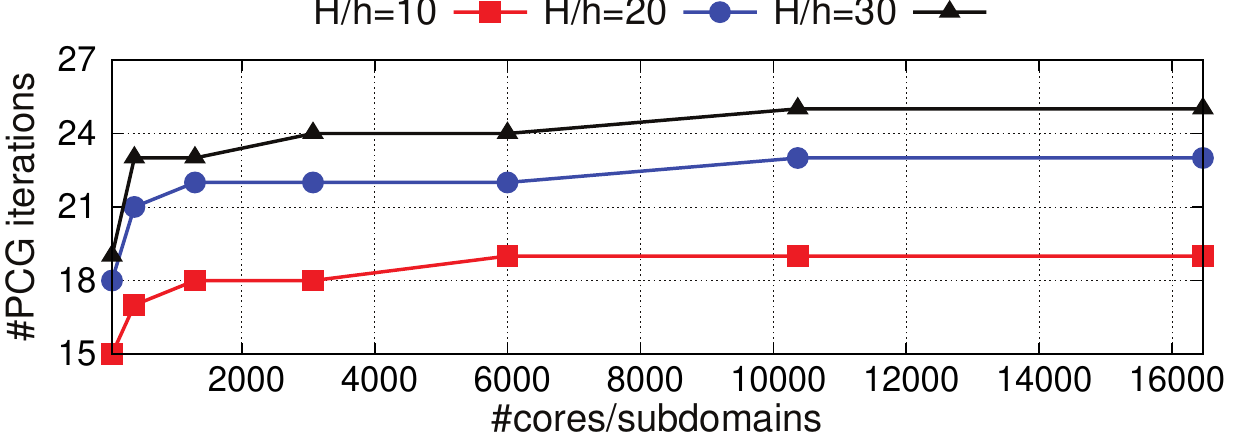}
      \caption{\#iterations}
      \label{fig-homoit}
      \vspace*{0.5cm}
  \end{subfigure}
  \begin{subfigure}[b]{0.45\textwidth}
    \centering
    \includegraphics[width=\textwidth]{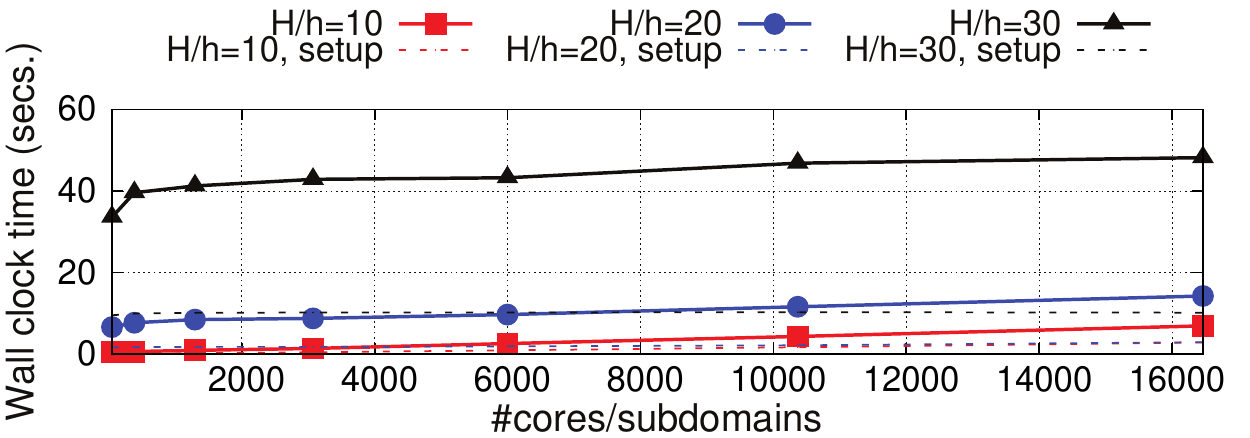}
    \caption{Preconditioner set-up + solve}
    \label{fig-homowt} 
    \vspace*{0.5cm}
  \end{subfigure}

  \begin{subfigure}[b]{0.45\textwidth}
    \includegraphics[width=\textwidth]{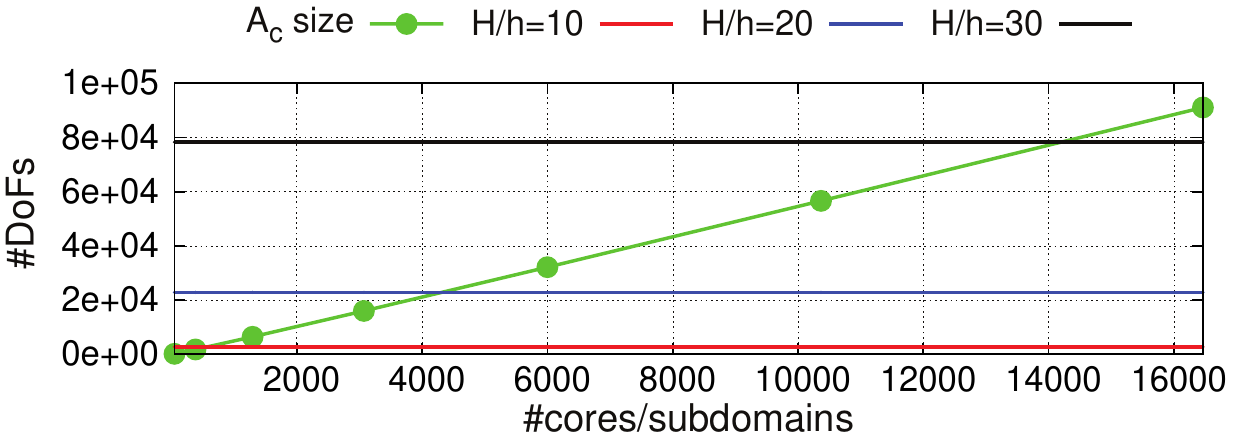}
    \caption{Coarse \REV{problem size and subdomain problem sizes.}}
    \label{fig-homosizes} 
  \end{subfigure}
  \begin{subfigure}[b]{0.45\textwidth}
      \includegraphics[width=\textwidth]{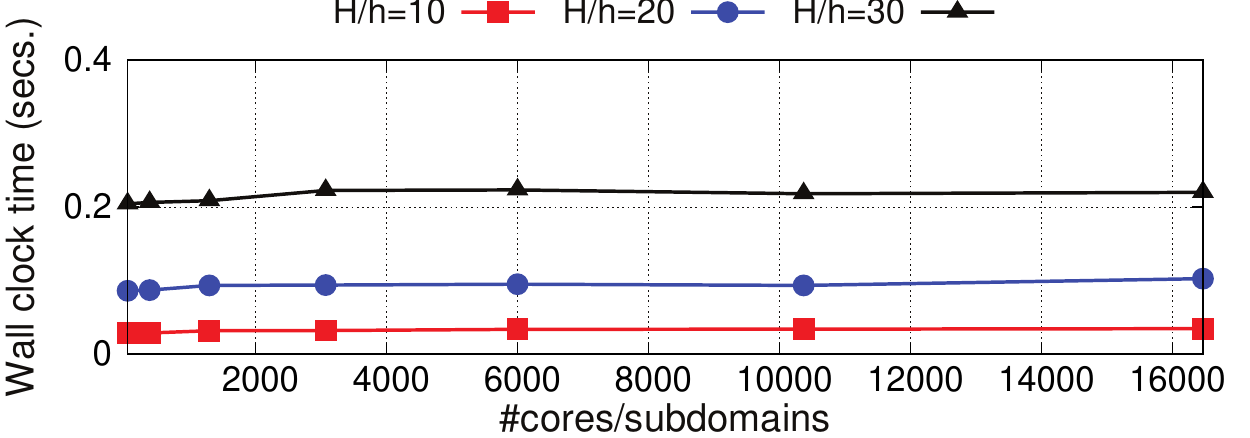}
    \caption{Change of basis set-up times}
    \label{fig-homocb}
  \end{subfigure}
  \caption{Weak scalability results for first order edge \ac{fe} with a constant distribution of materials for different local problem $\frac{H}{h}$ sizes.
           \REV{Subdomain problem sizes in \ref{fig-homosizes} are given for comparison purposes against the coarse problem size.}}
  \label{fig-homo_weak}
\end{figure}

Let us first consider homogeneous coefficients $\curlc = \massc = 1.0$ for the whole domain $\Omega$. We test the problem with different local problem sizes $\frac{H}{h}$ and \ac{fe} orders. In this case, the \ac{pbbddc} preconditioner reduces to the standard \ac{bddc} preconditioner since $\partition = \pbpartition$. In  \fig{fig-homo_weak}, we present weak scalability results for the homogeneous problem up to 16464 subdomains with different local problem sizes $\frac{H}{h}=\{10,20,30\}$, where the largest case has more than \REV{$10^9$ \acp{dof}}. We present the number of solver iterations until convergence in \fig{fig-homoit}, and employed wall clock times in \fig{fig-homowt}, which are composed by the preconditioner set-up time and the solution time with the \ac{bddc} preconditioned solver. The plots indicate that both the algorithm and its implementation in \FEMPAR have excellent weak scalability properties. Provided that the algorithm overlaps fine and coarse tasks, coarse tasks computing times are masked as long as they not exceed computing times for local problems, which allow us to observe excellent weak scalability times for the largest case $\frac{H}{h}=30$ in \fig{fig-homowt}. The size of both local and coarse problems is presented in \fig{fig-homosizes}. Finally, we add a plot (\fig{fig-homocb}) of the time needed to set-up the change of basis, which includes the edge partition algorithm (Alg. \ref{alg-edge_part}) to detect problematic cases, to show that consumed time is only dependent on the local problem size and not significant compared to the one spent in the solver run. 

\begin{figure}[ht!]
    \begin{subfigure}[b]{0.45\textwidth}
      \centering
      \includegraphics[width=\textwidth]{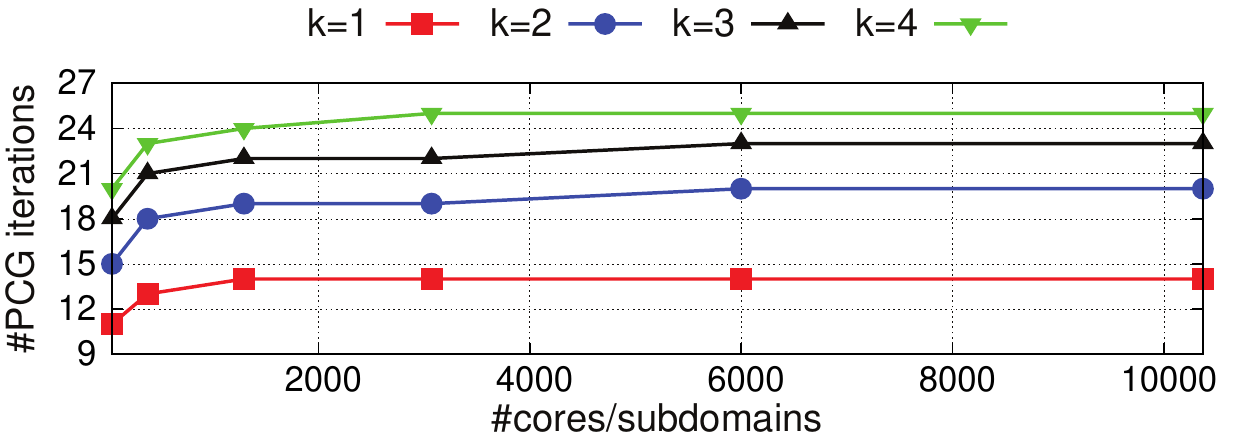}
      \caption{\#iterations}
      \label{fig-homord_it}
  \end{subfigure}
  \begin{subfigure}[b]{0.45\textwidth}
    \centering
    \includegraphics[width=\textwidth]{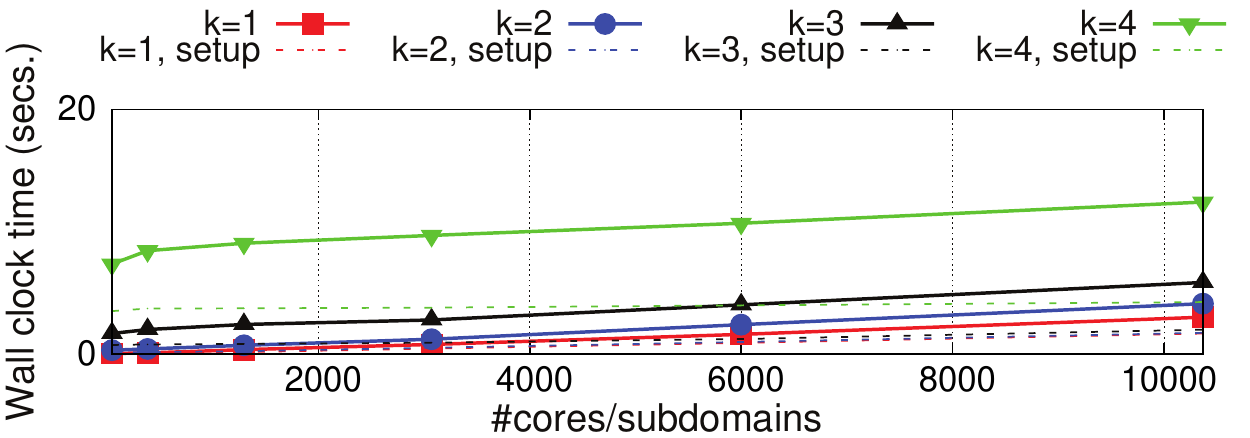}
    \caption{Preconditioner set-up + solve}
    \label{fig-homord_wt}
  \end{subfigure} 
  
  \begin{subfigure}[b]{0.48\textwidth}
    \includegraphics[width=\textwidth]{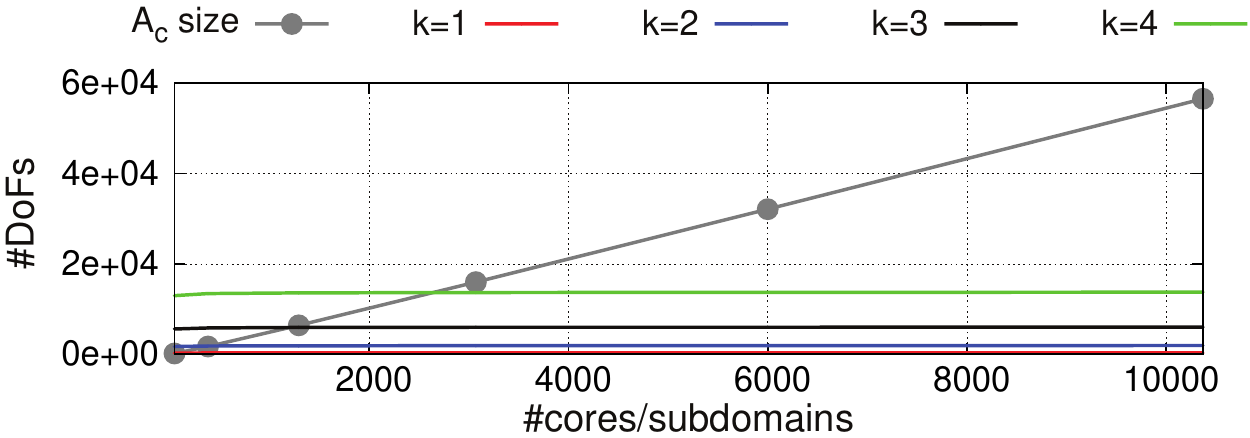}
    \caption{Coarse \REV{problem size and subdomain problem sizes.}}
    \label{fig-homord_sizes}
  \end{subfigure}
  \begin{subfigure}[b]{0.45\textwidth}
      \includegraphics[width=\textwidth]{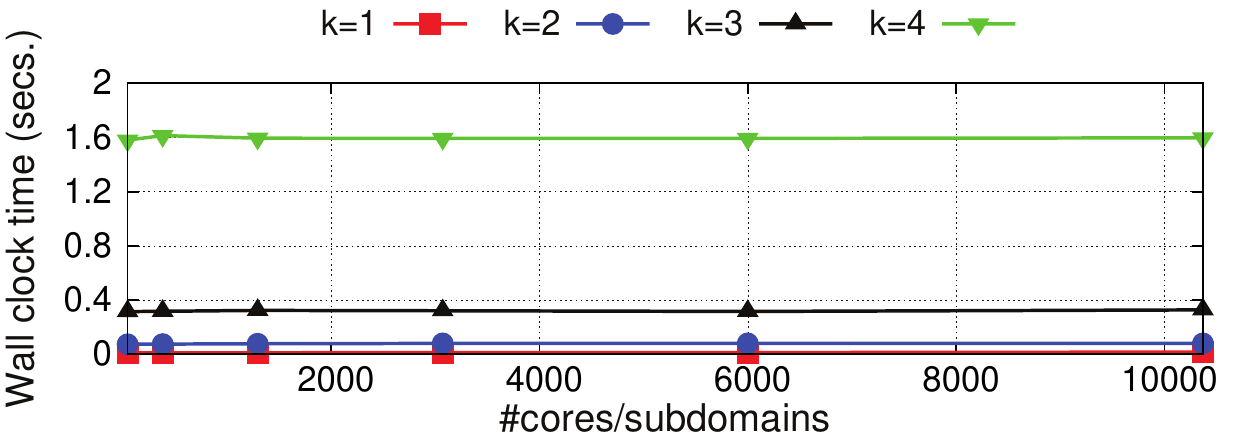}
    \caption{Change of basis set-up times}
    \label{fig-homord_cb}
  \end{subfigure}
  \caption{Weak scalability results for different order edge \ac{fe} with a constant distribution of materials and a local problem $\frac{H}{h}=10$. 
           \REV{Subdomain problem sizes in \ref{fig-homord_sizes} are given for comparison purposes against the coarse problem size.}}
  \label{fig-ho_weak}
\end{figure}

\fig{fig-ho_weak} shows the weak scalability results for an homogeneous problem with constant coefficients and different \ac{fe} orders up to 4. First, the number of iterations is (asymptotically) constant, thus the method is scalable. In this case, local \REV{problem sizes} are such that the coarse problem is larger from an \REV{small} number of subdomains (see \fig{fig-homord_sizes}), thus it is reflected in the solver times plot in \fig{fig-homord_wt}. In \fig{fig-homord_cb}, the time spent in the set-up of the change of basis is presented. Out of the presented results for the homogeneous case, a clear conclusion can be drawn for the standard \ac{bddc}: the algorithm and its implementation have excellent weak scalability properties.  

\subsection{Multi-material problem}

\begin{figure}[t!]
    \centering
    \begin{subfigure}[t]{0.35\textwidth}
        \includegraphics[width=\textwidth]{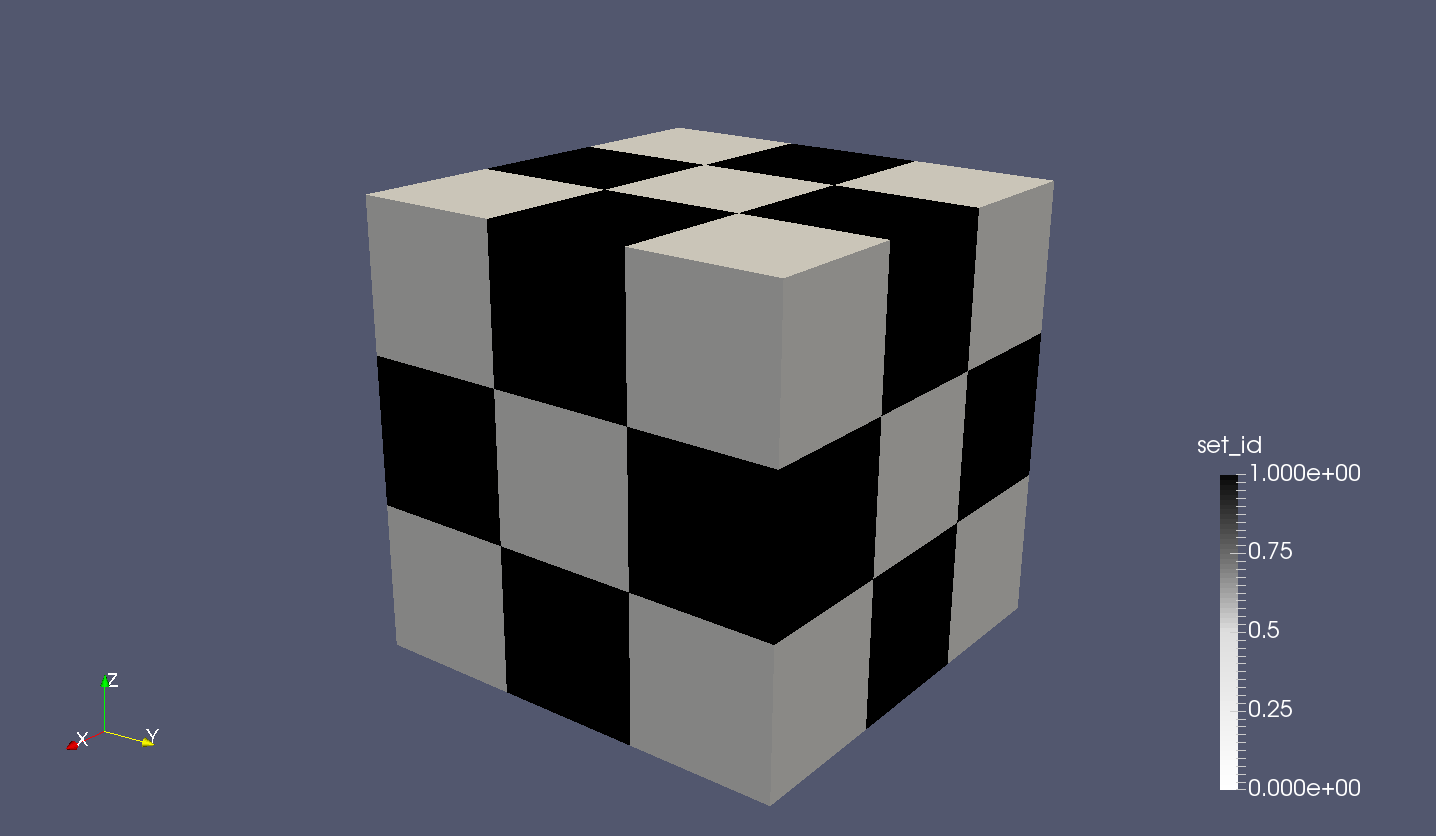}
        \caption{ Checkerboard distribution of coefficients, aligned with the partition and thus $\pbpartition = \partition$. }
        \label{fig-check}
    \end{subfigure} \hspace{1cm}
    \begin{subfigure}[t]{0.25\textwidth}
        \includegraphics[width=\textwidth]{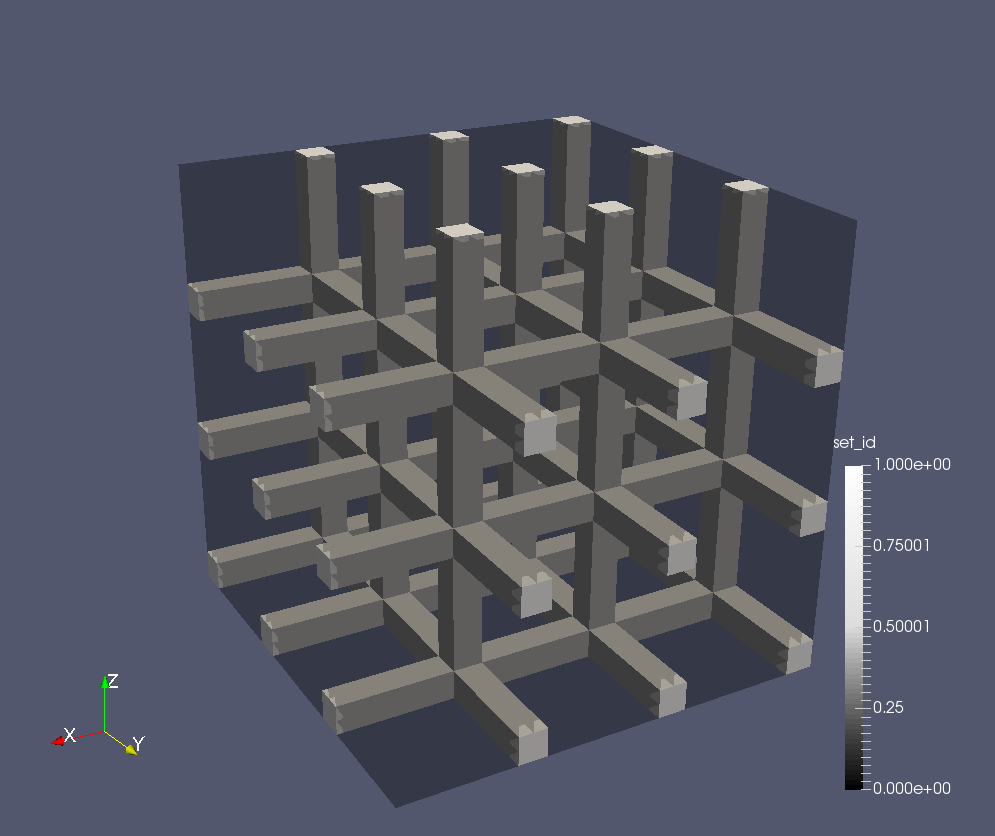}
        \caption{ Channel distribution of coefficients with $N=3$ and $\gamma=0.2$. }
        \label{fig-chan}
    \end{subfigure} 
    \caption{Bi-material distribution cases in a structured $3\times 3\times 3$ partition of the unit cube $[0,1]^3$.}\label{fig-mcube}
\end{figure}

\subsubsection{Checkerboard distribution}\label{subsec-chekd}
The checkerboard arrangement of coefficients is a widely used distribution of materials to test the robustness of the \ac{bddc} algorithms for problems in $H$(curl) against the jump of coefficients across the interface \cite{toselli_dual-primal_2006, dohrmann_bddc_2016}. In short, it is a bi-material distribution of subdomain-wise constant coefficients such that every subdomain presents a jump of coefficients through the faces to all its neighbours. For the sake of ease, let us distinguish between black and white subdomain materials, see \fig{fig-check}. Note that in the checkerboard distribution case, the jumps of coefficients are aligned with the partition, thus $\pbpartition = \partition$.

\begin{figure}[ht!]
    \centering
    \includegraphics[width=0.8\textwidth]{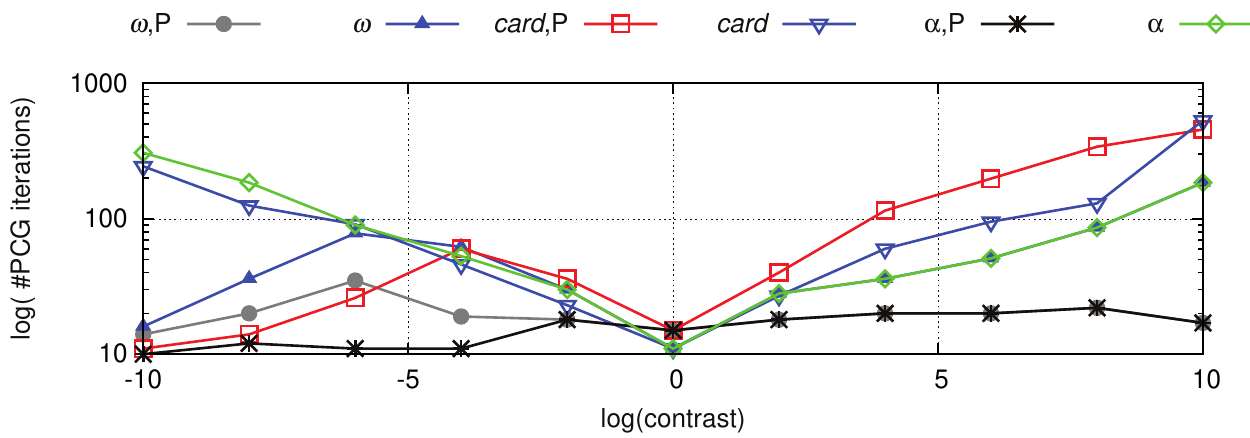}
  \caption{Number of iterations for first order edge \ac{fe} with a $3\times 3 \times 3$ partition of the unit cube and $H/h=8$. A checkerboard arrangement of materials is defined: $\curlc_{\rm white} = 1.0$,  $\massc_{\rm white} = 1.0$ and $\curlc_{\rm black} = 10^i$ and $\massc_{\rm black} = 10^{-i}$, leading to a contrast $=\frac{\curlc_{\rm black}}{\massc_{\rm black}}=\frac{10^i}{10^{-i}} = 10^{2i}$. Labels include scaling information, where P denotes perturbation of the preconditioner. }
  \label{fig-contrast_chk}
\end{figure}

We first test the robustness of the algorithm against the contrast of the coefficients.  
Consider a $3\times 3 \times 3$ partition of a unit cube domain with a checkerboard distribution of coefficients such that $\curlc_{\rm white} = \massc_{\rm white} = 1.0$ and $\curlc_{\rm black} = 10^i$, $\massc_{\rm black} = 10^{-i}$. The contrast is defined here as $\frac{\curlc_{\rm black}}{\massc_{\rm black}}$. With the variation of the value for $i$ in the range $[-5,5]$ we test all the possible scenarios, namely the mass dominated problem $(i<0)$ and the curl dominated problem $(i>0)$. The number of iterations with the contrast of the coefficients for different configurations of the preconditioner is presented in \fig{fig-contrast_chk}. Out of the plot, the most salient property is the robustness of the perturbed preconditioner (see \sect{subsec-pertur}) with the contrast of the coefficients. In fact, the original formulation of the preconditioner suffers from a large number of iterations when the contrast between the two coefficients is large, specially in the curl-dominated case. Therefore, the proposed perturbation of the preconditioner is essential to achieve a robust preconditioner, in the case where both coefficients $\curlc$ and $\massc$ jump across the interface.  
Clearly, the perturbed formulation only has a (negligible) negative impact in the case $i=0$, since actually no jump occurs across the interface. In the curl-dominated limit, $\curlc$ and $\omega$-based scalings show the same behaviour, as it is suggested by the definition of $\omega$ when $\curlc >> \massc$. On the other hand, when the coefficient $\massc$ becomes dominant, the choice of \emph{cardinal} and $\omega$-based scalings also leads to good scalability results in the limit. In \REV{summary}, the combination of the perturbed formulation and $\curlc$-based scaling is the most robust approach. Unless otherwise stated, this combination will be used throughout the section. 

Let us now consider a checkerboard arrangement of coefficients such that $\alpha_{\rm white} = 10^2$,  $\massc_{\rm white} = 1.0$ and $\curlc_{\rm black} = 10^4$ and $\massc_{\rm black} = 10^{-2}$. In order to show the importance of the perturbed formulation of the preconditioner for jumps of both coefficients across interfaces, we collect the number of iterations for the original and perturbed preconditioner in Tabs. \ref{tab-ws_original} and \ref{tab-ws_perturbed}, respectively. The problem is solved with a $P=N \times N \times N$ partition of the unit cube and the $\omega$-based scaling is employed in both cases. Iteration counts for the perturbed preconditioner are noticeably lower in all cases without exception. 

\begin{table}
\centering
\begin{subtable}{.5\textwidth}
\centering
\begin{tabular}{lrrrrrrr}\hline
  $\boldsymbol{\frac{H}{h}}$ / \bf{P}       &  \bf{$2^3$}  & \bf{$3^3$}   & \bf{$4^3$}  & \bf{$5^3$} & \bf{$6^3$} & \bf{$7^3$} & \bf{$8^3$}\\ \hline
         \bf{4}      &  14      &    24    &  35      &  38      &   40    & 40 & 41      \\ 
         \bf{8}      &  26      &    37    &  61      &  65      &   70    & 69   & 70        \\ 
         \bf{12}     &  31      &    52    &  72      &  78      &   82    & 82   & 84        \\ \hline
\end{tabular}
      \caption{Standard \ac{bddc} preconditioner}
      \label{tab-ws_original}
\end{subtable}
\begin{subtable}{.5\textwidth}
\centering
\begin{tabular}{lrrrrrrr}\hline
   $\boldsymbol{\frac{H}{h}}$ / \bf{P}        &  \bf{$2^3$}  & \bf{$3^3$}   & \bf{$4^3$}  & \bf{$5^3$} & \bf{$6^3$} & \bf{$7^3$} & \bf{$8^3$}\\ \hline
         \bf{4}      &  8      &    9    &  10      &  10      &   11    & 12     & 12      \\ 
         \bf{8}      &  12      &   14    &  16      &  16      &   17    & 17    & 17        \\ 
         \bf{12}     &  15      &    22    &  21      &  21      &   21    & 21   & 21        \\ \hline
\end{tabular}
        \caption{Perturbed \ac{bddc} preconditioner}
        \label{tab-ws_perturbed}
\end{subtable}

\caption{Weak scalability in terms of number of iterations for both preconditioners. Checkerboard distribution of materials with $\alpha_{\rm white} = 10^2$,  $\massc_{\rm white} = 1.0$ and $\curlc_{\rm black} = 10^4$ and $\massc_{\rm black} = 10^{-2}$. }
\end{table}

\begin{figure}[ht!]
    \begin{subfigure}[t]{0.48\textwidth}
      \centering
      \includegraphics[width=\textwidth]{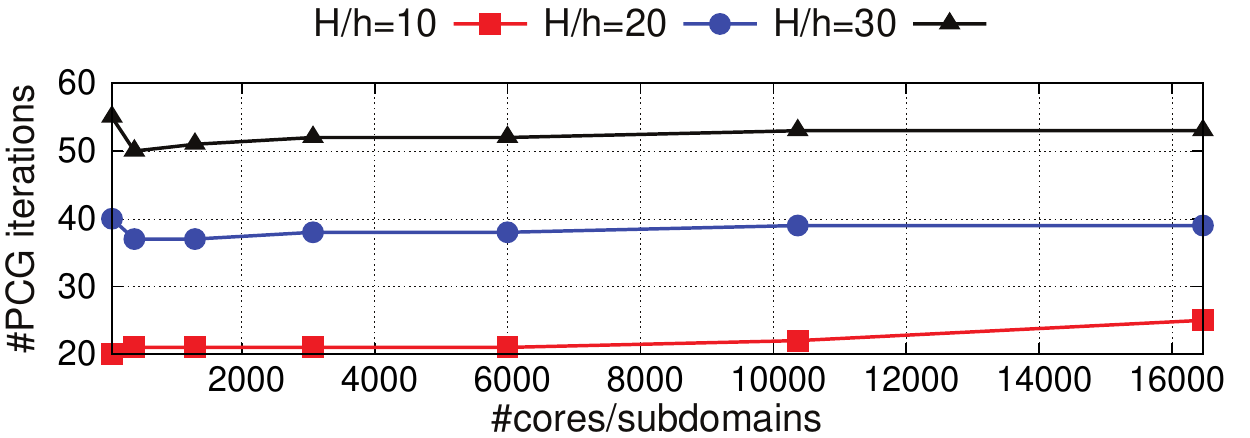}
      \caption{\#iterations}
      \label{fig-ck_iter}
  \end{subfigure}
  \begin{subfigure}[t]{0.48\textwidth}
    \centering
    \includegraphics[width=\textwidth]{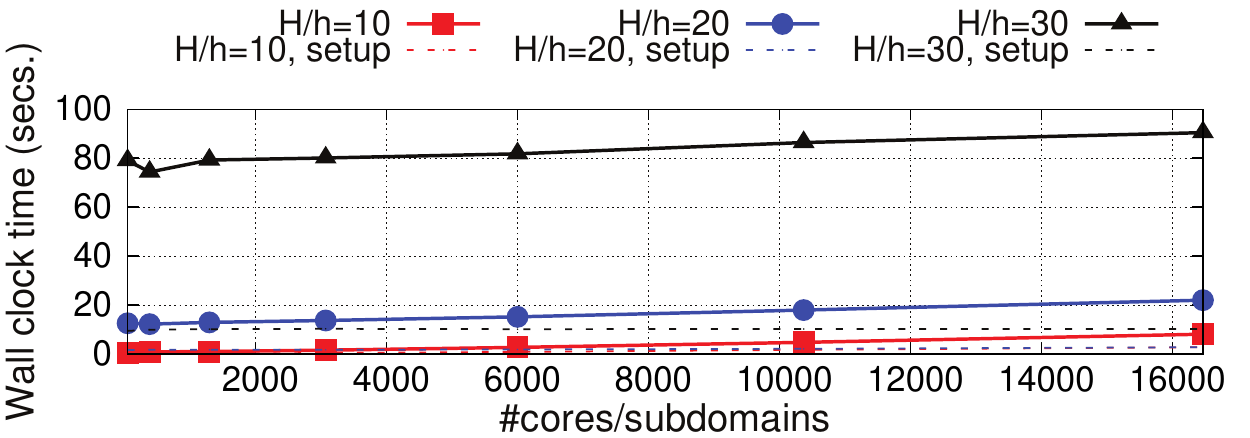}
    \caption{Preconditioner set-up + solve}
  \end{subfigure} 

  \caption{Weak scalability results for first order edge \ac{fe} with a checkerboard distribution of materials: $\curlc_{\rm white}=10^2$,  $\massc_{\rm white} = 1.0$ and $\curlc_{\rm black} = 10^4$ and $\massc_{\rm black} = 10^{-2}$.}
  \label{fig-checks_weak}
\end{figure}

\begin{figure}[ht!]
    \begin{subfigure}[t]{0.48\textwidth}
      \centering
      \includegraphics[width=\textwidth]{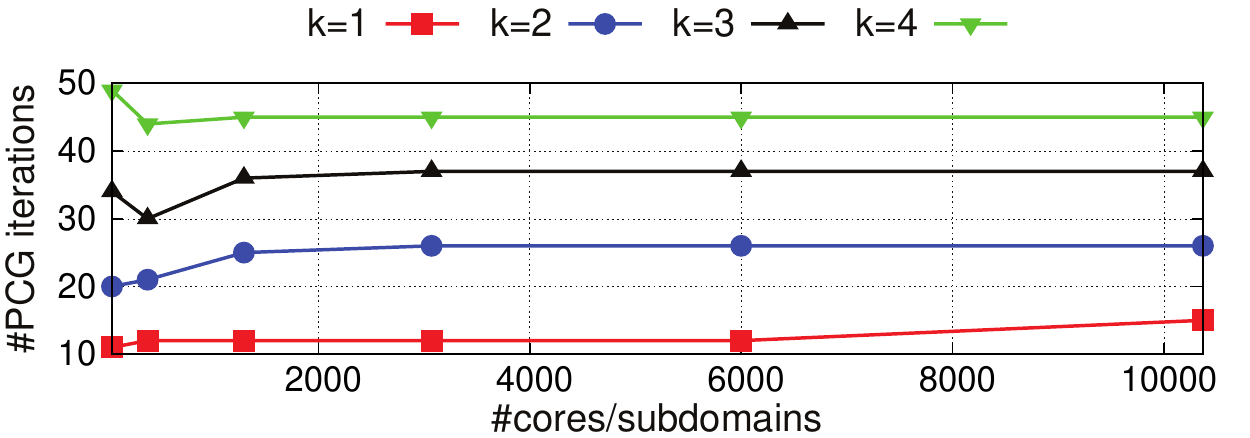}
      \caption{\#iterations}
      \label{fig-ck_iters_order}
  \end{subfigure}
  \begin{subfigure}[t]{0.48\textwidth}
    \centering
    \includegraphics[width=\textwidth]{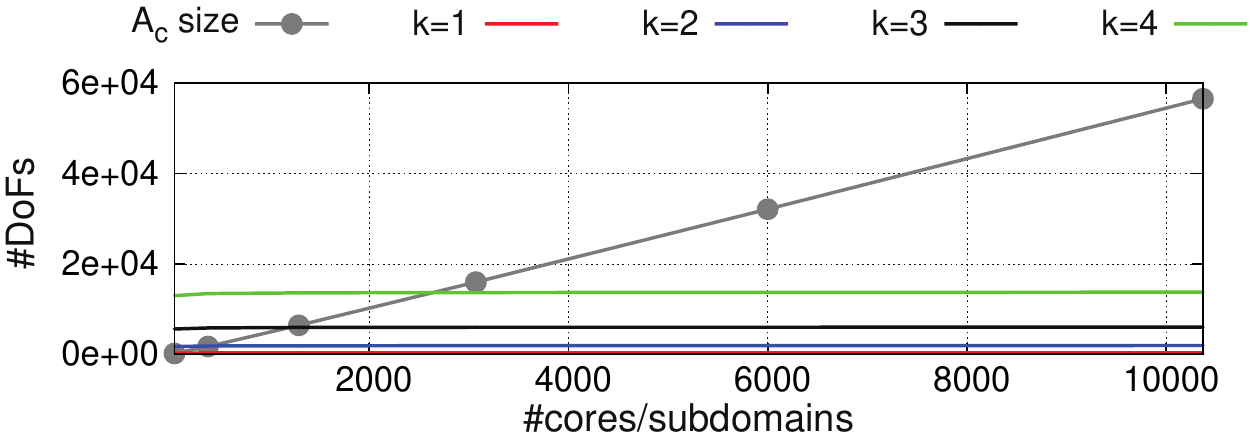}
    \caption{Problem sizes}
  \end{subfigure} 

  \caption{Weak scalability results for different order edge \acp{fe} and $H/h=4$ with a checkerboard distribution of materials: $\curlc_{\rm white}=10^2$,  $\massc_{\rm white} = 1.0$ and $\curlc_{\rm black} = 10^4$ and $\massc_{\rm black} = 10^{-2}$.}
  \label{fig-checks_weak_order}
\end{figure}

Once we have shown the importance of the perturbed formulation, we present a weak scalability analysis up to 16,464 subdomains and the checkerboard arrangement of materials with the perturbed preconditioner. Problem sizes in this experiment coincide to the ones presented for the homogeneous problem in \fig{fig-homosizes}. As expected, plots in \fig{fig-checks_weak} show excellent scalability properties of the preconditioner in this case, i.e., the preconditioner is robust with jumps of coefficients across the interface. Although higher values of $\frac{H}{h}$ lead to a significantly higher number or iterations, these ones are (asymptotically) constant and remain in a reasonable range, see \fig{fig-ck_iter}. 
On the other hand, \fig{fig-ck_iters_order} presents the number of iterations for different order \acp{fe} and problem size $H/h=4$, which also is shown to be scalable. Out of the contrast and scalability results, we would like to remark the following issues. First, the perturbed formulation of the preconditioner is essential to achieve a robust preconditioner. Second, the method is weakly scalable for problems with high coefficient jumps across interfaces for different local sizes and \ac{fe} orders. 

\begin{figure}[t!]
    \centering
    \begin{subfigure}[t]{0.4\textwidth}
        \includegraphics[width=\textwidth]{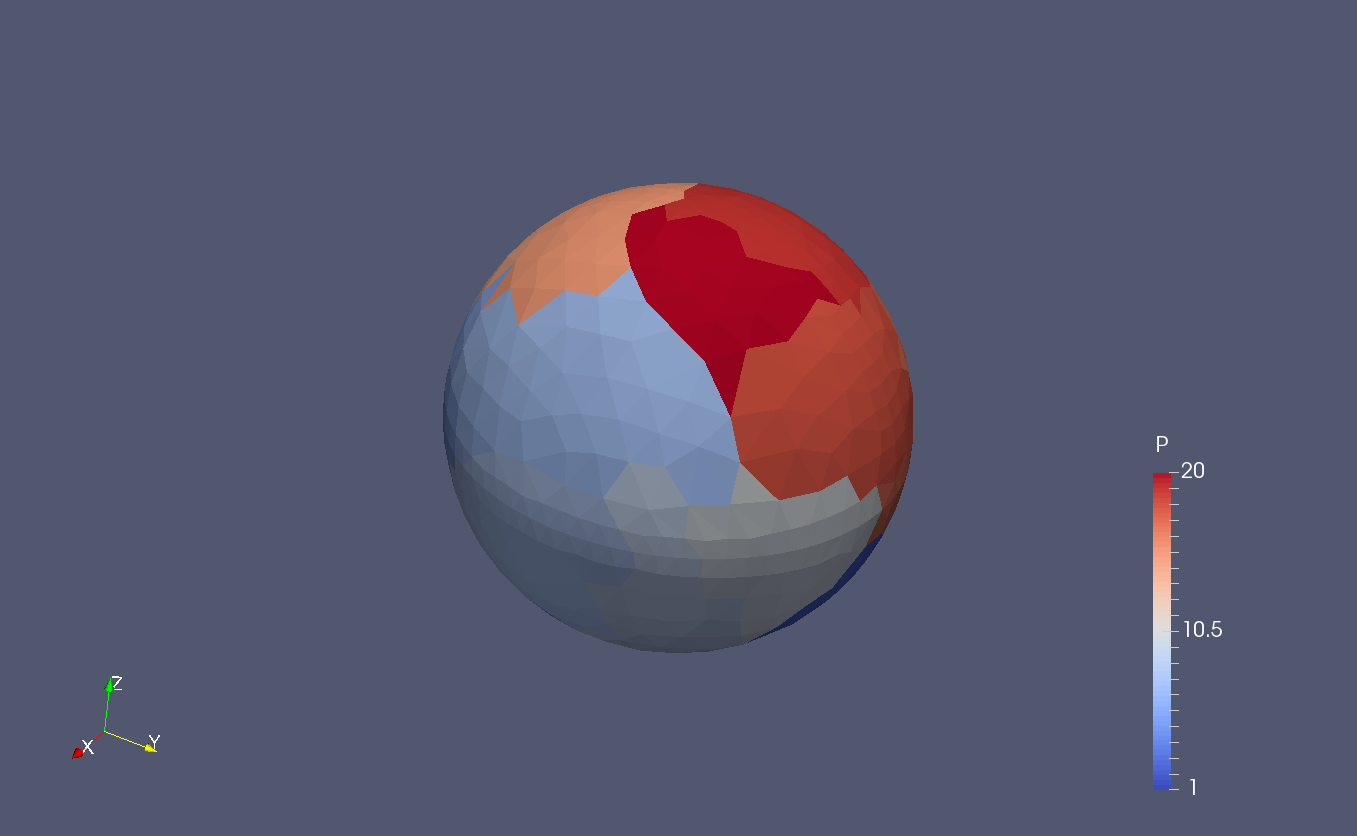}
        \caption{ Partition of the sphere into 20 subdomains. }
        \label{fig-sphere_part}
    \end{subfigure} \hspace{1cm}
    \begin{subfigure}[t]{0.4\textwidth}
        \includegraphics[width=\textwidth]{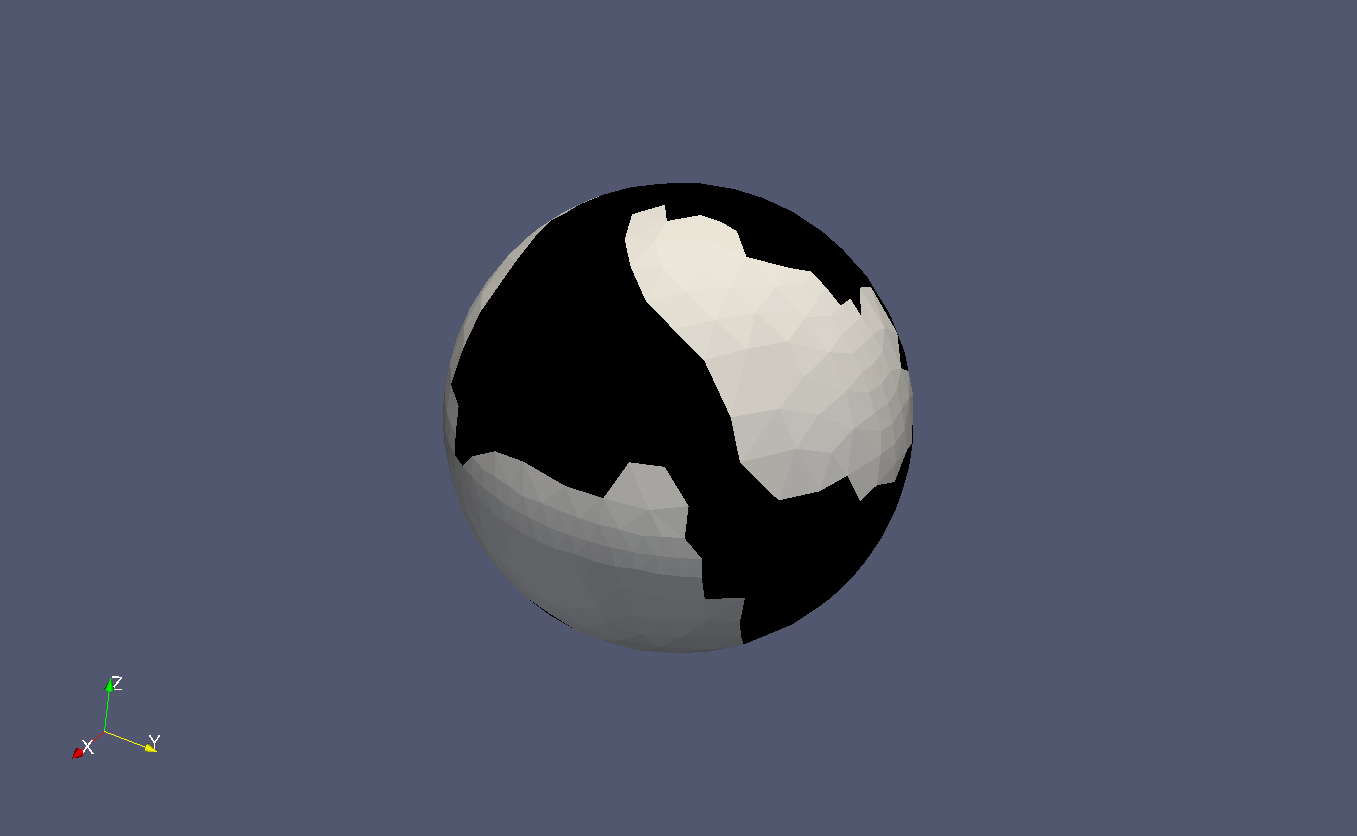}
        \caption{ Bi-material random distribution of materials with constant properties on each subdomain. }
        \label{fig-sphere_ids}
    \end{subfigure} 
    \caption{Sphere partition and distribution of materials.}\label{fig-sphere}
\end{figure}

\begin{figure}[t!]
    \centering
        \includegraphics[width=0.8\textwidth]{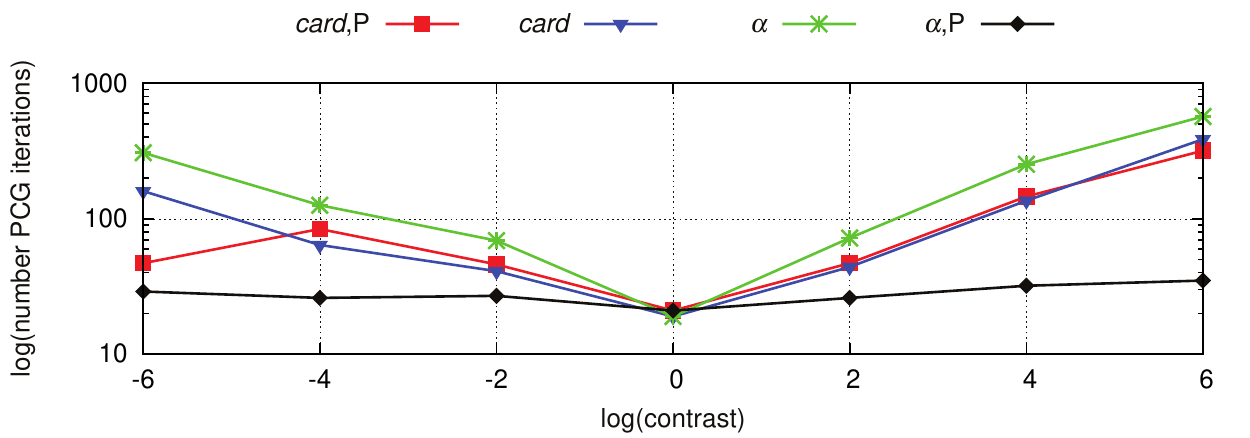}
        \caption{ Robustness for a tetrahedral mesh for different scalings. Material parameters defined as $\curlc_{\rm white}=10^i$, $\curlc_{\rm black}=1.0$, $\massc_{\rm white}=10^{-i}$ and $\massc_{\rm black}=1.0$, contrast defined $\frac{\curlc_{\rm white}}{\massc_{\rm white}}$. Mesh partitioned into 20 subdomains and random assignment of materials. Labels include scaling information, where P denotes perturbation of the preconditioner.}
        \label{fig-sphere_contrast}
\end{figure}

In order to show the robustness of the method not only with structured, regular hexahedral meshes we solve the problem for a spherical domain and partition with a graph partitioner METIS. Let us consider a spherical domain with $R=0.5$, discretized with an unstructured tetrahedral mesh containing around 50,000 cells. In order to achieve high contrast of coefficients across interfaces, a bi-material distribution of coefficients is assigned such white or black subdomain-wise constant materials are randomly assigned, see \fig{fig-sphere_ids} for an illustration. The definition of the sets of coefficients is $\curlc_{\rm white}=10^2$,  $\massc_{\rm white} = 1.0$ and $\curlc_{\rm black} = 10^4$ and $\massc_{\rm black} = 10^{-2}$. \fig{fig-sphere_contrast} shows the number of iterations with the original and the perturbed formulation of the preconditioner. The perturbed preconditioner, combined with a $\curlc$-based scaling, is the unique method shown to be robust with regard to the coefficients contrast, reproducing the behaviour observed in the structured case. 

\subsubsection{Multiple channels}

In this arrangement of materials we have a background domain, denoted by black region, and a set of inclusions, denoted by white regions, that cross the domain from one boundary to the opposite one in parallel to the axes directions. We include one channel per direction per subdomain so that with an increasing number of $P$ subdomains we are solving a harder problem with $P$ channels. Channels are parallel to the axes and are positioned in the lowest (i.e, minimum $x,y,z$ coordinates) corners within every subdomain. They have a squared cross-section of size $\gamma H$, thus occupying a $\gamma^2 | \tilde \Omega_i | $ volume in every subdomain, see \fig{fig-chan} for an illustration of a $3\times3\times3$ partition of the unit cube with the described channel inclusions. The distribution of coefficients is such that contains coefficient jumps within each subdomain and also across all interfaces, thus the perturbed formulation of the \ac{pbbddc} preconditioner will be employed. 

Let us first compare the number of iterations for the \ac{pbbddc} preconditioner against the ones that one would have with the standard \ac{bddc}, where the definition of \emph{globs} is generated with the original partition $\partition$. In \sect{subsec-chekd}, the effectiveness of the perturbation formulation has been empirically shown. Consequently, in order to provide a fair comparison among them, the perturbed formulation is considered for both preconditioners. On the other hand, while $\curlc$-based scaling is shown to be the most robust approach for \ac{pbbddc} preconditioner, it miserably fails when considered with the standard \emph{globs}, i.e., given by $\partition$. In this case, a better result is obtained with \emph{cardinality} scaling. Tabs. \ref{tab-bddc_comparison} and \ref{tab-pbbddc_comparison} show iteration counts for the described \ac{bddc} and \ac{pbbddc} preconditioners, respectively, for the solution of the channels problem with $\gamma=0.5$ and a partition of the unit cube into $P=3\times3\times3$ subdomains with local size $H/h=8$. We define the coefficients $\curlc_{\rm black} = \massc_{\rm black}=1.0$, while we distinguish between $\curlc_{\rm white}=10^i$ and $\massc_{\rm white}=10^{-i}$, which allow us to define the contrast as $\frac{\curlc_{\rm white}}{\massc_{\rm white}}$. As expected, the \ac{pbbddc} preconditioner is robust with the contrast of coefficients. On the other hand, the number of iterations increases for the \ac{bddc} preconditioner in the curl-dominated case. 
 
\begin{table}
\centering
\begin{subtable}{.5\textwidth}
\centering

\begin{tabular}{lrrrrr}\hline
  $\boldsymbol{\frac{H}{h}}$/\bf{c} &  $10^{-4}$  & $10^{-2}$ &  1.0 & $10^2$ & $10^4$ \\ \hline
         \bf{4}      &  36      &    29    &  13      &  31      &   74      \\ 
         \bf{8}      &  67      &    36    &  16      &  38      &   104      \\  \hline
\end{tabular}
\caption{\ac{bddc} preconditioner}
\label{tab-bddc_comparison}
\end{subtable}
\begin{subtable}{.5\textwidth}
\centering
\begin{tabular}{lrrrrr}\hline
  $\boldsymbol{\frac{H}{h}}$/\bf{c} &  $10^{-4}$  & $10^{-2}$ &  1.0 & $10^2$ & $10^4$ \\ \hline
         \bf{4}      &  14      &    14    &  11      &  13      &   14      \\ 
         \bf{8}      &  18      &    19    &  16      &  17      &   20      \\  \hline
\end{tabular}
\caption{\ac{pbbddc} preconditioner}
\label{tab-pbbddc_comparison}
\end{subtable}

\caption{Comparison in number of iterations for both preconditioners in a $3\times 3\times 3$ partition. Channel distribution of materials with $\gamma=0.5$ and $\curlc_{\rm black} = \massc_{\rm black}=1.0$, $\curlc_{\rm white}=10^i$ and $\massc_{\rm white}=10^{-i}$. Contrast defined as $c=\frac{\curlc_{\rm white}}{\massc_{\rm white}}$. }
\end{table}

\begin{figure}[ht!]
    \begin{subfigure}[b]{0.48\textwidth}
      \centering
      \includegraphics[width=\textwidth]{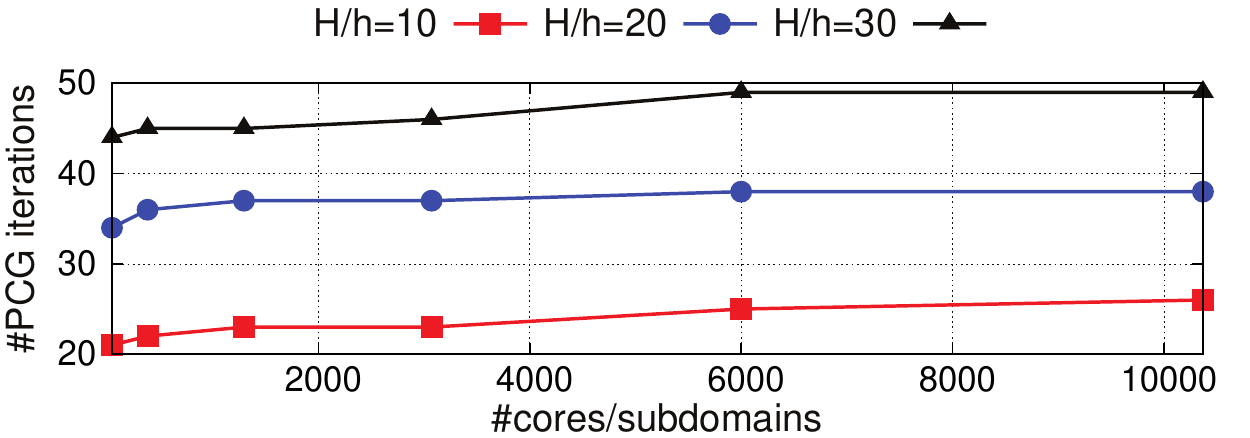}
      \caption{\#iterations}
      \label{fig-cn_iters_Hh}
  \end{subfigure}
  \begin{subfigure}[b]{0.48\textwidth}
    \centering
    \includegraphics[width=\textwidth]{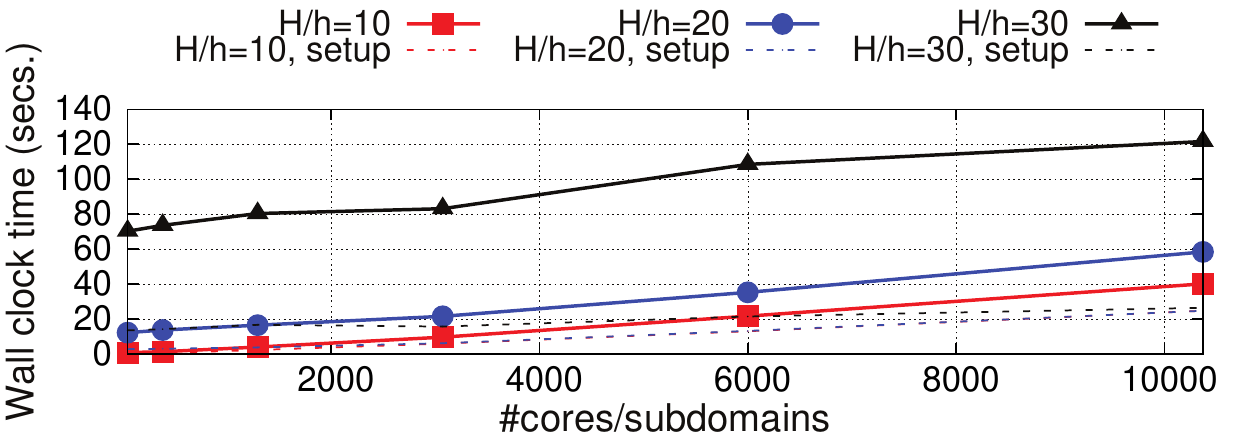}
    \caption{Problem sizes}
    \label{fig-cn_total_times_Hh}
  \end{subfigure} 
  
    \begin{subfigure}[b]{0.48\textwidth}
    \centering
    \includegraphics[width=\textwidth]{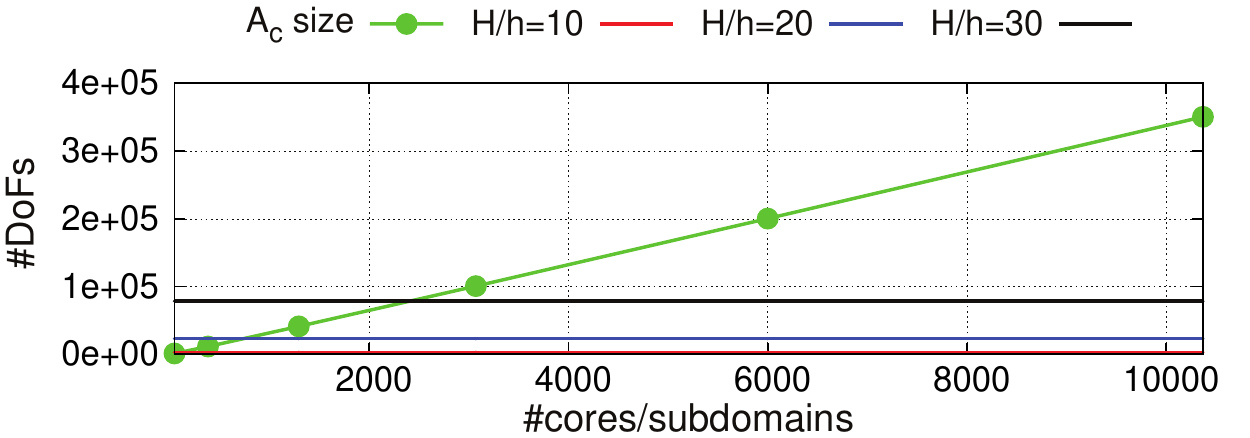}
    \caption{Problem sizes}
    \label{fig-cn_sizes_Hh}
  \end{subfigure} 
  \begin{subfigure}[b]{0.48\textwidth}
      \centering
      \includegraphics[width=\textwidth]{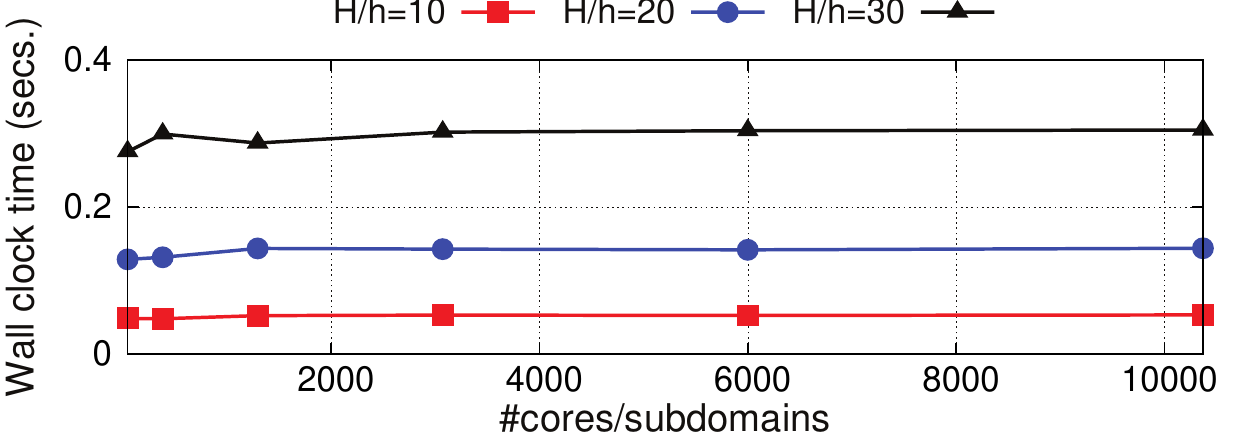}
      \caption{\REV{Change of basis set-up time}}
      \label{fig-cn_cb_times_Hh}
  \end{subfigure}

  \caption{Weak scalability results for first order edge \ac{fe} with a channel distribution of materials: $\curlc_{\rm white} = 10^2$,  $\massc_{\rm white} = 1.0$ and $\curlc_{\rm black} = 10^4$ and $\massc_{\rm black} = 10^{-2}$.}
  \label{fig-chan_weak}
\end{figure}
\begin{figure}[ht!]
    \begin{subfigure}[b]{0.48\textwidth}
      \centering
      \includegraphics[width=\textwidth]{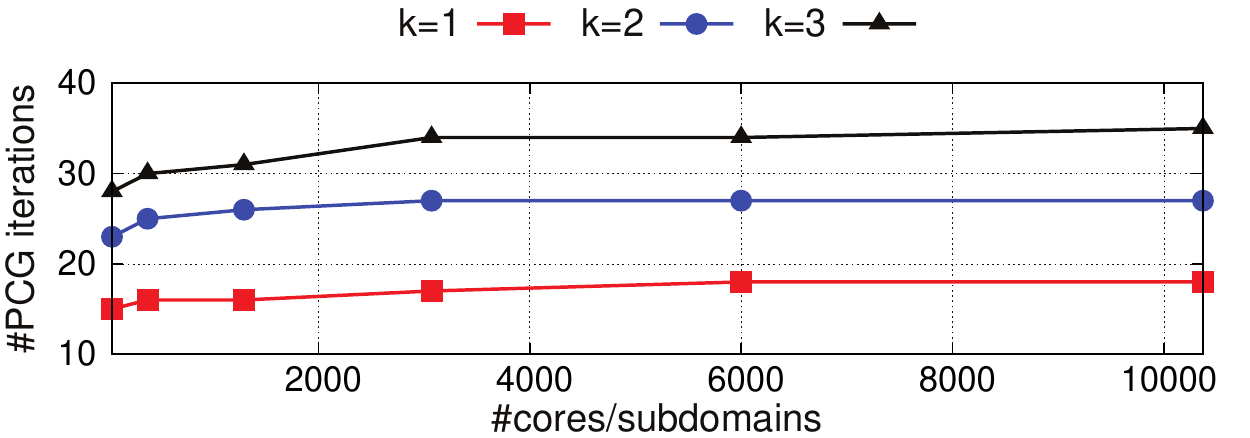}
      \caption{\#iterations}
      \label{fig-cn_iters_order}
  \end{subfigure}
  \begin{subfigure}[b]{0.48\textwidth}
    \centering
    \includegraphics[width=\textwidth]{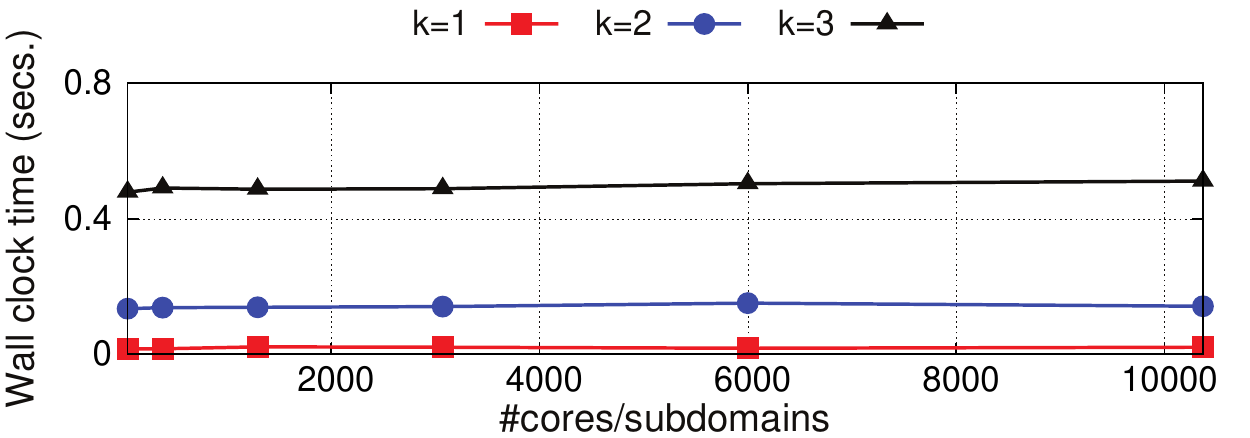}
    \caption{Change basis set-up time}
    \label{fig-cb_cb_times_order} 
  \end{subfigure} 

  \caption{Weak scalability results for different order edge \ac{fe} and $H/h=4$ with a channel distribution of materials: $\curlc_{\rm white} = 10^2$,  $\massc_{\rm white} = 1.0$ and $\curlc_{\rm black} = 10^4$ and $\massc_{\rm black} = 10^{-2}$.}
  \label{fig-chan_weak_order}
\end{figure}

The following experiment evaluates the weak scalability properties for a channel-type distribution of materials. In Figs. \ref{fig-chan_weak} and \ref{fig-chan_weak_order} we present weak scalability results for different problem sizes and \ac{fe} orders. The most salient property out of \REV{these} plots is that the number of iterations is asymptotically constant for all cases. However, coarse problem sizes become larger as the partition into \ac{pb}-subdomains generates a higher number of coarse \acp{dof}, see \fig{fig-cn_sizes_Hh}. In this context, the coarse problem is larger than local \REV{problem sizes} from an \REV{small} number of subdomains, thus coarse tasks will predominate computing times precluding wall clock time scalability, as it is shown in \fig{fig-cn_total_times_Hh}. 
In Figs. \ref{fig-cn_cb_times_Hh} and \ref{fig-cb_cb_times_order} we present scalable wall clock times for the change of basis set-up, for different local problem sizes and \ac{fe} orders. 

We would like to remark that the proposed \ac{pbbddc} preconditioner is weakly scalable for the number of iterations until convergence not only with regard to the jump of coefficients across interfaces but also for distributions of different materials within each subdomain. A multilevel version of the preconditioner for curl-conforming spaces \cite{Zampini2017}, not addressed in this work, is expected to push forward the limits of the computing times scalability results.
 
\subsection{Heterogeneous problems}
In this section we study the scalability of the \ac{rpbbddc} method for problems where the coefficients $\curlc, \massc$ are described by continuous (at least element-wise) functions, which contain high contrasts for their maximum and minimum values. \REV{In order to build the rPB-partition, the approach based on the aggregation of cells with their coefficients on the same interval \eq{eq-interv} is used, see \sect{subsec-rpb_imp}.  }

\subsubsection{Periodic analytical functions}
\begin{figure}[ht!]
    \begin{subfigure}[t]{0.47\textwidth}
      \centering
      \includegraphics[width=\textwidth]{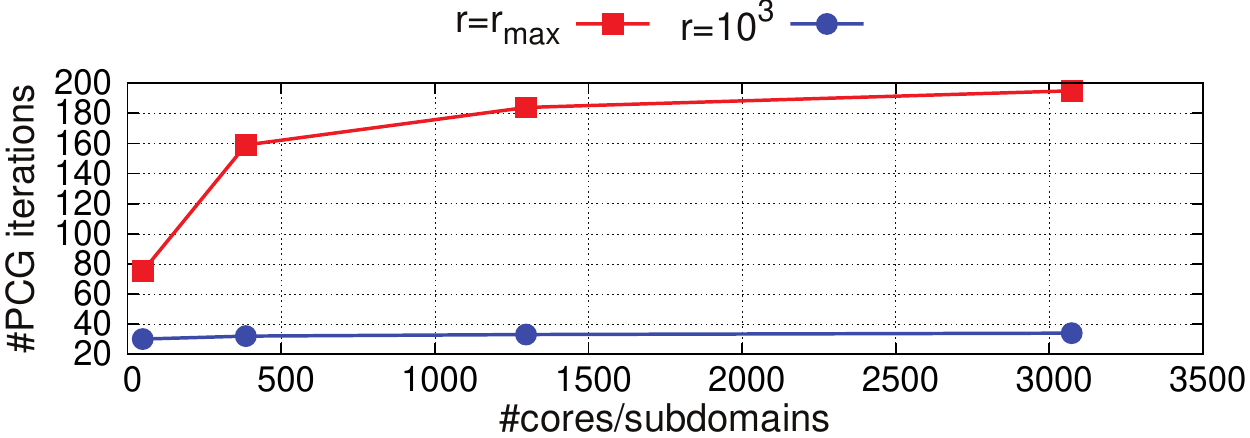}
      \caption{Number of iterations}
      \label{fig-het_i_rho} 
  \end{subfigure}
  \hspace{0.2cm}
    \begin{subfigure}[t]{0.5\textwidth}
    \centering
    \includegraphics[width=\textwidth]{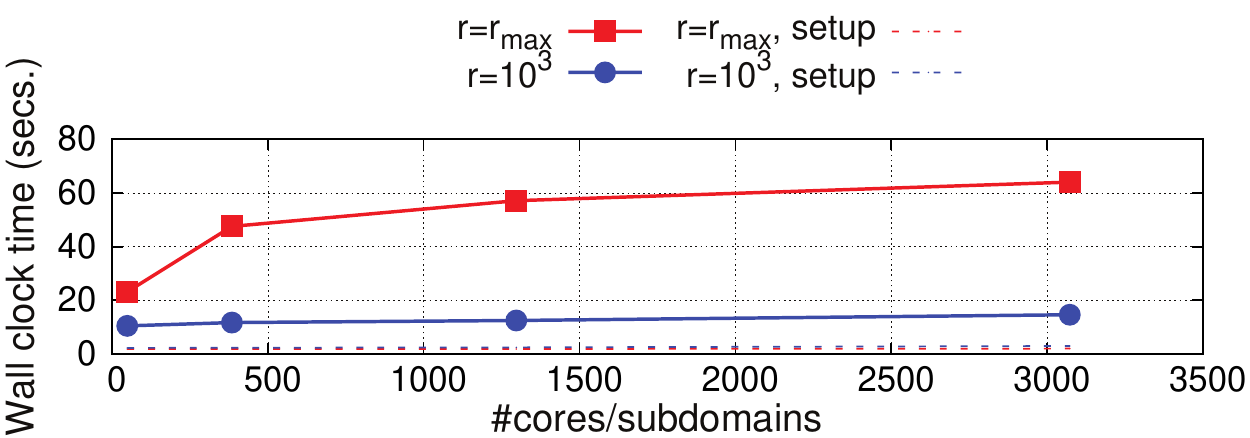}
    \caption{Set-up and total solver time.}
    \label{fig-het_t_rho}
  \end{subfigure} 
  \caption{Weak scalability for the \ac{rpbbddc} when only an heterogeneous $\curlc$ is considered, $\massc=1.0$.}
  \label{fig-rpb_rho}
\end{figure}

\begin{figure}[ht!]
    \begin{subfigure}[t]{0.47\textwidth}
      \centering
      \includegraphics[width=\textwidth]{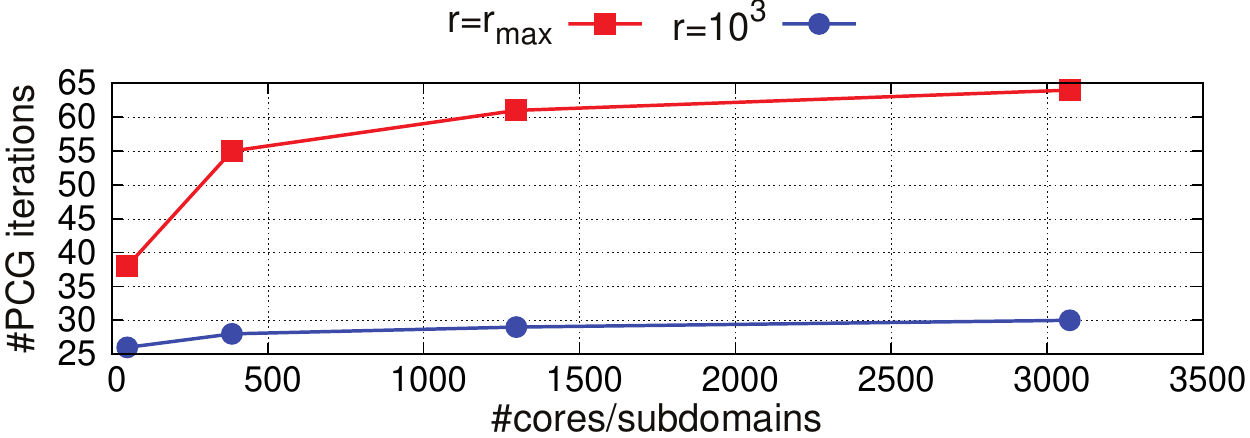}
      \caption{Number of iterations}
      \label{fig-het_i_mu} 
  \end{subfigure}
  \hspace{0.2cm}
    \begin{subfigure}[t]{0.5\textwidth}
    \centering
    \includegraphics[width=\textwidth]{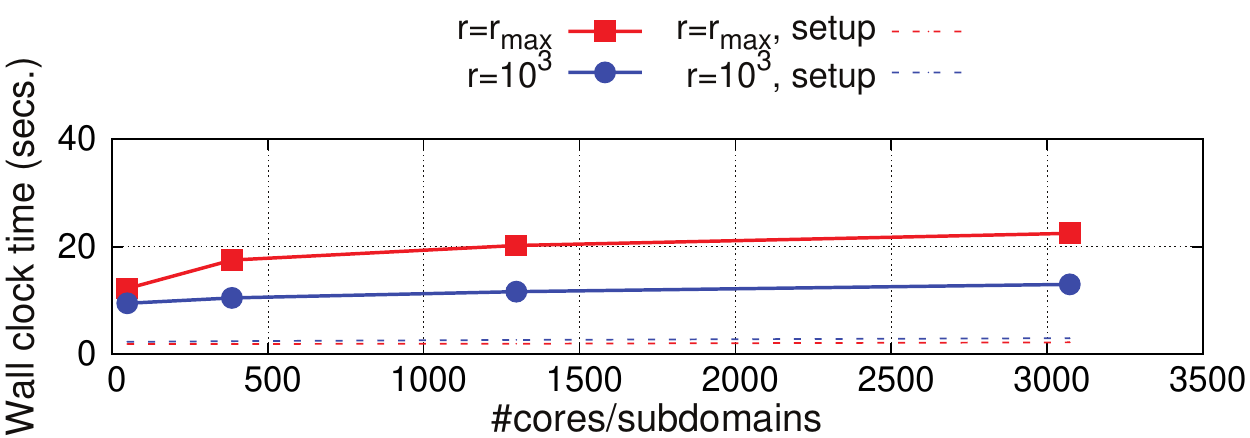}
    \caption{Set-up and total solver time.}
    \label{fig-het_t_mu}
  \end{subfigure} 

  \caption{Weak scalability for the \ac{rpbbddc} when only an heterogeneous $\massc$ is considered, $\curlc=1.0$.}
  \label{fig-rpb_mu}
\end{figure}

\begin{figure}[ht!]
    \begin{subfigure}[t]{0.47\textwidth}
      \centering
      \includegraphics[width=\textwidth]{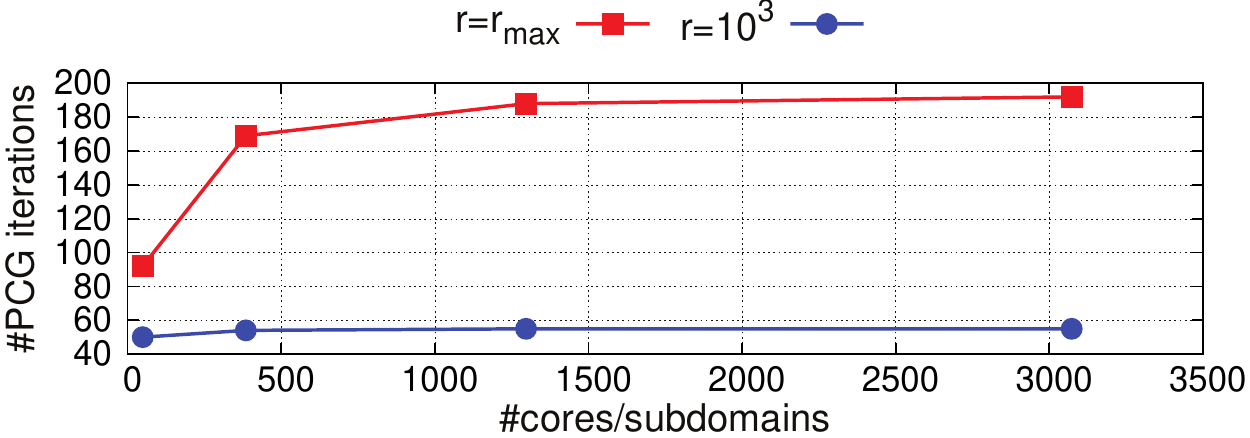}
      \caption{Number of iterations}
      \label{fig-het_i_both} 
  \end{subfigure}
  \hspace{0.2cm}
    \begin{subfigure}[t]{0.5\textwidth}
    \centering
    \includegraphics[width=\textwidth]{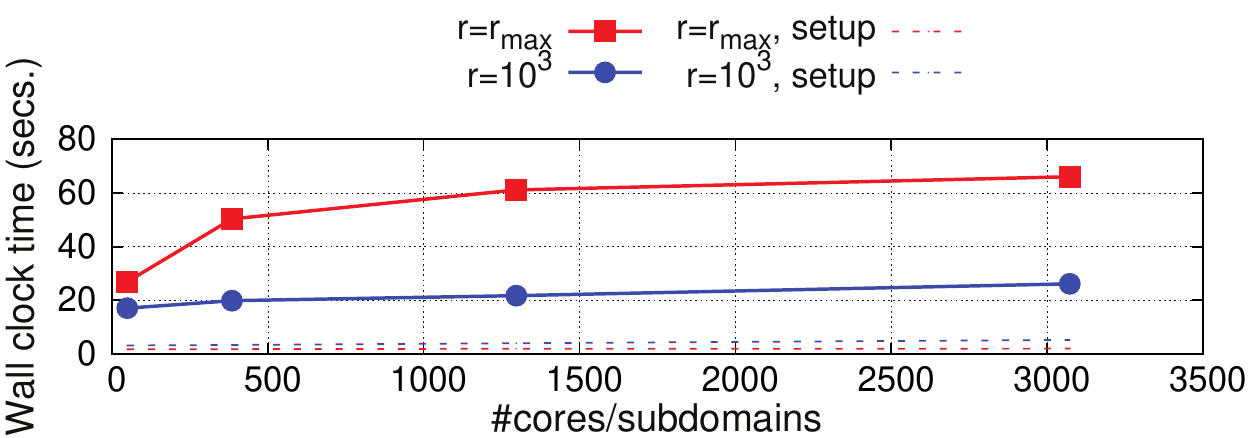}
    \caption{Set-up and total solver time.}
    \label{fig-het_t_both}
  \end{subfigure} 
  \caption{Weak scalability for the \ac{rpbbddc} when both coefficients are heterogeneous.}
  \label{fig-rpb_both}
\end{figure}

In this case, $\curlc$ and $\massc$ are defined as exponential functions with a sinusoidal exponent such that the function is periodic on the domain and the number of peaks scales with the number of original subdomains in $\partition$, thus solving a harder problem as we increase the number of processors. In particular, let us consider $\log(\curlc) = \frac{c_{\rm max}}{2} \sin(N_x \pi x)$ and $\log(\massc) = \frac{c_{\rm max}}{2} \sin(N_y \pi y)$, where $N_x,N_y$ denotes the number of subdomains per ($x,y$) direction in a $P = N_x \times N_y \times N_z$ structured partition, (see $\massc$ depicted in \fig{fig-beta_fun} for the case $c_{\rm max}=6$, $N_x=N_y=3$). Clearly, the maximum contrast within each coefficient is given by $r_{\rm max} = 10^{c_{\rm max}}$. We present weak scalability results up to 3072 subdomains for two different thresholds $r=\{ r_{\rm max}, 10^3 \}$, and for a local problem size of $\frac{H}{h}=20$ in three different scenarios: coefficients $\curlc$ (\fig{fig-rpb_rho}), $\massc$ (\fig{fig-rpb_mu}) or both are heterogeneous (\fig{fig-rpb_both}), being set to $\curlc=1.0$, $\massc=1.0$ otherwise. Out of these plots, we can draw some conclusions. First, the case where only $\massc$ is heterogeneous converges in a lower number of iterations compared to problems with heterogeneous $\curlc$. Secondly, the consideration of lower values for $r$ consequently results in larger coarse problems, but its size is only (approximately) doubled when only one coefficient is heterogeneous 
or (approximately) quadrupled when both are defined heterogeneous. 
In fact, in the range of subdomains considered in this experiment, the coarse problem computational times in all cases can be masked by local problem ones (Figs. \ref{fig-het_t_rho}, \ref{fig-het_t_mu} and \ref{fig-het_t_both}). 
Finally, and most salient, the \ac{rpbbddc} method with $r=10^3$ is weakly scalable in all cases with an excellent reduction in the number of iterations and computing times compared to the case where $r=r_{\rm max}$. 

\subsubsection{High Temperature Superconductors}
\begin{figure}[ht!]
    \begin{subfigure}[t]{0.31\textwidth}
      \centering
      \includegraphics[width=\textwidth]{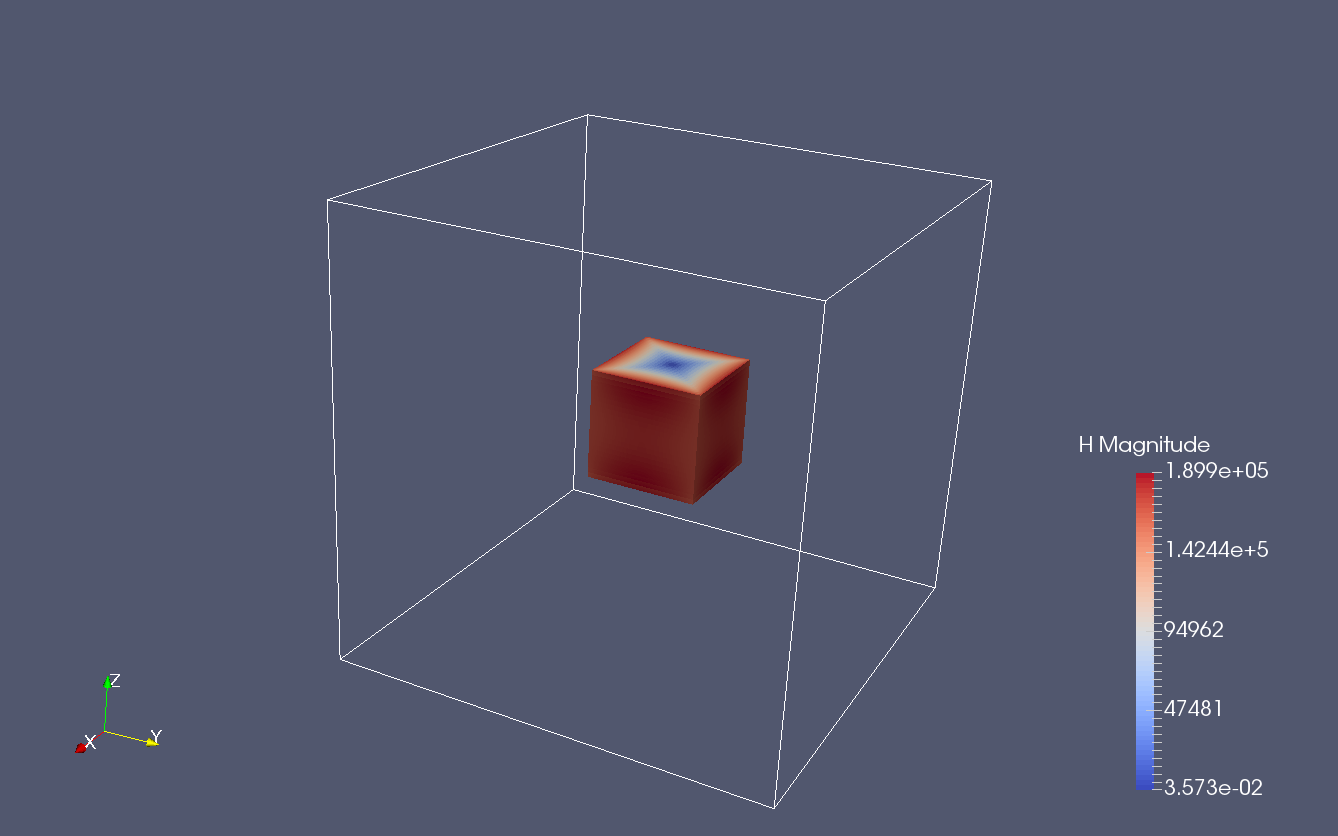}
      \caption{Magnetic field strength in the \ac{hts} device surrounded by a dielectric box, for which only the outline is depicted.}
      \label{fig-hts_domain} 
  \end{subfigure}
  \hspace{0.2cm}
  \begin{subfigure}[t]{0.31\textwidth}
    \centering
    \includegraphics[width=\textwidth]{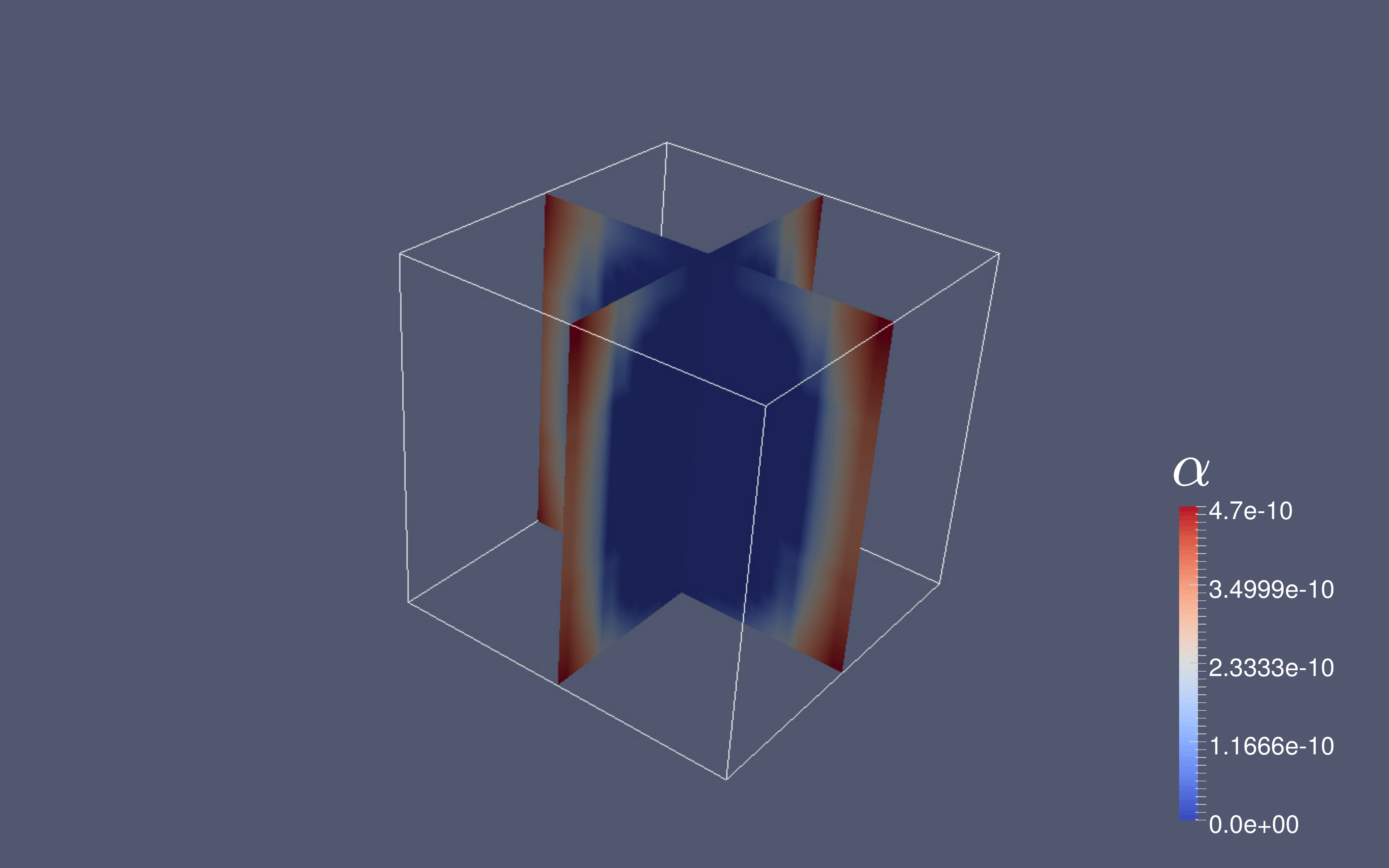}
    \caption{Distribution of $\curlc$ in the \ac{hts} device.}
    \label{fig-alpha_hts}
  \end{subfigure} 
  \hspace{0.2cm}
    \begin{subfigure}[t]{0.31\textwidth}
    \centering
    \includegraphics[width=\textwidth]{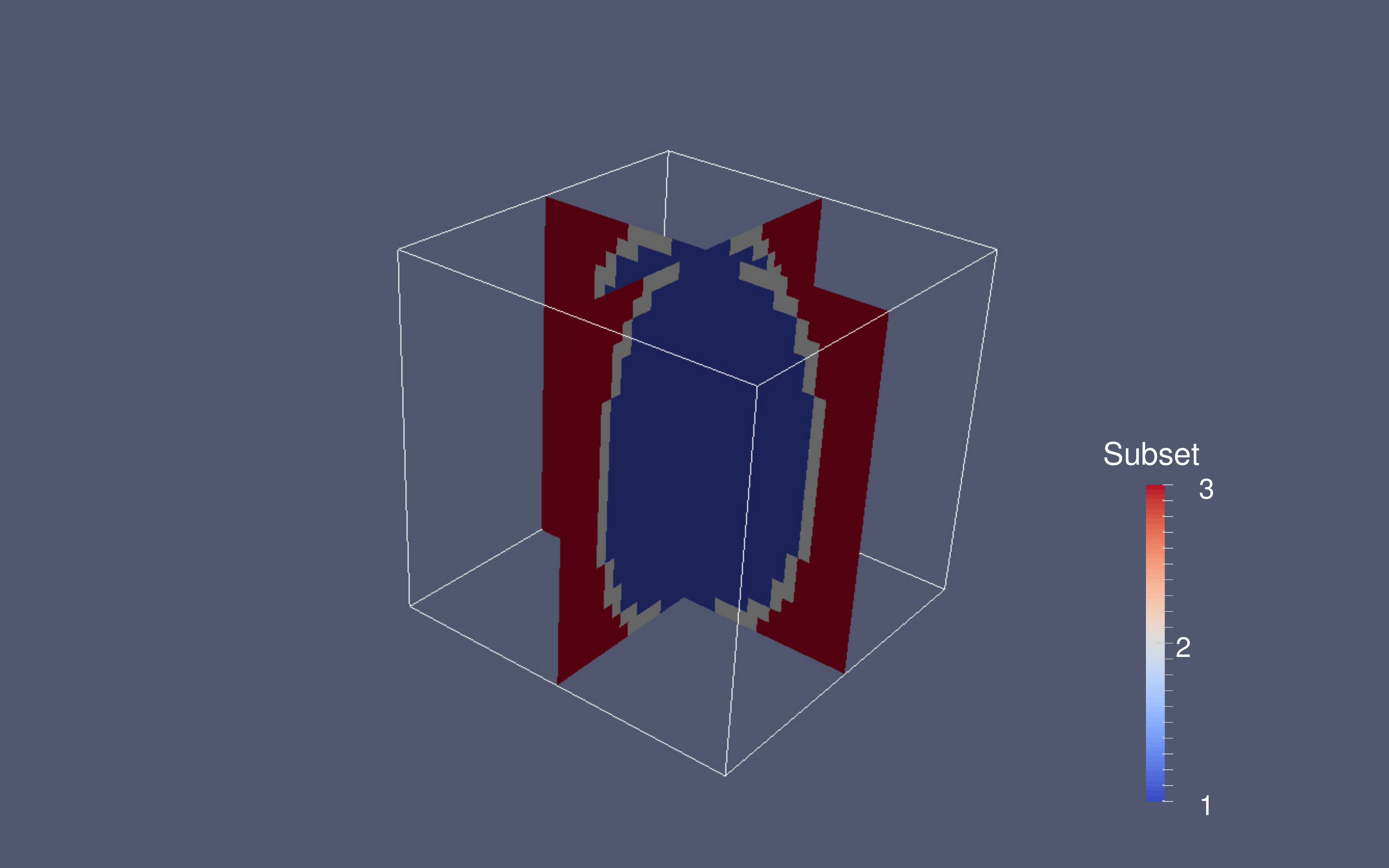}
    \caption{Subsets arising by the \ac{rpb}-partition with $r=10^2$ in the \ac{hts} device.}
    \label{fig-subsets_hts}
  \end{subfigure} 

  \caption{Domain and \ac{hts} device for $t=4$~ms.}
  \label{fig-hts_example}
\end{figure}

Next, we study the scalability of the algorithm with a practical application, the modelling of \ac{hts}. The problem consists in the magnetization of a superconducting cube completely  surrounded by a dielectric material (see \fig{fig-hts_domain}), subjected to an external AC magnetic field. The formulation \eq{eq-strong_maxwell} arises in the time-domain quasi-static approximation of the Maxwell's equations for solving the magnetic field, see \cite{molm_hts} for details. Furthermore, the standard Backward Euler method is used to perform time integration over a time interval $[0,T]$, so let us define a time partition $\{0=t^0, t^1, \ldots, t^N=T \}$ into $N$ time elements. Then, the form \eq{eq-strong_maxwell} can be used to compute the magnetic field for a particular time $t^n$, provided the solution on the previous time $\uc^{n-1}$. The coefficient $\massc$ is affected by the current time step size $\Delta t = (t^{n}-t^{n-1})$ as $\massc=\frac{\mu_0}{\Delta t}$, where $\mu_0 = 4\pi \cdot 10^{-7}$ is the magnetic permeability of the vacuum. While the dielectric material is modelled with a constant value for $\curlc = 10^{-3}$, the superconducting material behaviour is modelled with the stiff nonlinear dependence of the resistivity $\curlc$ with the solution as $\curlc= \curlc_0 \left( \frac{ \norm{\grad \times \uc}}{J_c} \right)^m$, with $m=100$, $J_c=10^{-8}$ and $\curlc_0=10^{-12}$. The equivalence with \eq{eq-strong_maxwell} is completed by considering the source term $\f = \massc \uc^{n-1}$ and the strong imposition of an external magnetic field $\uc^{n} \times \boldsymbol{n} = \uc_0^{n}$ over the whole boundary. For the time step $t^n$, the weak form of the nonlinear problem reads: find $\uc^{n} \in \Xg$ such that 
\begin{align}\label{eq-htsor}
(\curlc(\uc^n) \grad\times \uc^n, \grad \times \vc )  + \massc ( \uc^n, \vc ) = \massc ( \uc^{n-1}, \vc ) \quad \forall \vc \in \Xg.
\end{align}
In order to derive the linearized form with Newton's method we consider the current approximation $\uc^{n,k}$ and a (small) correction $\delta \uc^{n,k}$ for the iterate $k$ such that $\uc^{n,k+1} = \uc^{n,k} + \delta\uc^{n,k}$. We plug the expression in \eq{eq-htsor}, consider a first order Taylor expansion of $\curlc(\uc^{n,k+1})$ around $\uc^{n,k}$ and  neglect the quadratic terms with respect to $\delta \uc^{n,k}$, which yields the linearized problem: find $\delta \uc^{n,k} \in \Xg$ such that 

\begin{align}\label{eq-linearized_form}
\mathcal{J}( \uc^{n,k}, \delta \uc^{n,k}, \vc ) = -\mathcal{R} ( \uc^{n-1,k}, \uc^{n,k}, \vc ) \qquad \quad \forall \vc \in \Xg,  
\end{align} 
where 
\begin{subequations} 
\begin{align}
{\mathcal{J}( \uc^{n,k}, \delta \uc^{n,k}, \vc )} &= (\curlc(\uc^{n,k}) \grad\times \delta\uc^{n,k},\grad \times \vc ) + \massc ( \delta\uc^{n,k}, \vc ) + \\ &+  (\curlc'(\uc^{n,k}) \delta \uc^{n,k} \grad\times \uc^{n,k},\grad \times \vc ), \nonumber \\ 
{\mathcal{R}( \uc^{n-1,k}, \uc^{n,k}, \vc )} &= - \massc ( \uc^{n-1}, \vc ) + \massc ( \uc^{n,k}, \vc ) + (\curlc(\uc^{n,k}) \grad\times \uc^{n,k},\grad \times \vc ). \label{eq-reshts} 
\end{align}
\end{subequations}
Therefore, the \ac{rpbbddc} preconditioner is applied to the linearized problem \eq{eq-linearized_form} at every nonlinear iteration of every time step. We will focus on the performance of the linear solver, and the reader is directed to \cite{molm_hts} for a detailed exposition of the composition of the used transient nonlinear solver.

The problem is solved in $\Omega=[0,40]^3~{\rm mm}^3$, composed by an outer dielectric $\Omega_{\rm air}$ material which includes a concentric superconducting cube $\Omega_{\rm hts}$ of size 10~mm such that $\Omega = \Omega_{\rm hts} \cup \Omega_{\rm air}$, see \fig{fig-hts_domain}.   
There is no source term and Dirichlet-type boundary conditions are imposed over the entire boundary as the time-dependent magnetic field $\uc_{ext}=\frac{B_0}{\mu_0} [0,0,\sin(2\pi \omega t)]$, where $B_0 = 200~{\rm mT}$ and $\omega=50~{\rm Hz}$. We solve the problem in the time interval $[0,5]~{\rm ms}$, which corresponds to a quarter of a full cycle in the applied $\uc_{ext}$. Initial conditions are simply $\uc^0=\boldsymbol{0}$. The partition $\rpbpartition$ is obtained in all simulations for $r=10^2$. The nonlinear scheme is stopped when the $L_2$-norm of the nonlinear residual (\eq{eq-reshts}) is below $10^{-4}$, while the convergence criteria for the \ac{rpbbddc} preconditioned linear solver is the reduction of the initial $L_2$-norm of the residual of the linearized system by $10^{-8}$. 

\begin{figure}[ht!]
    \begin{subfigure}[t]{0.47\textwidth}
      \centering
      \includegraphics[width=\textwidth]{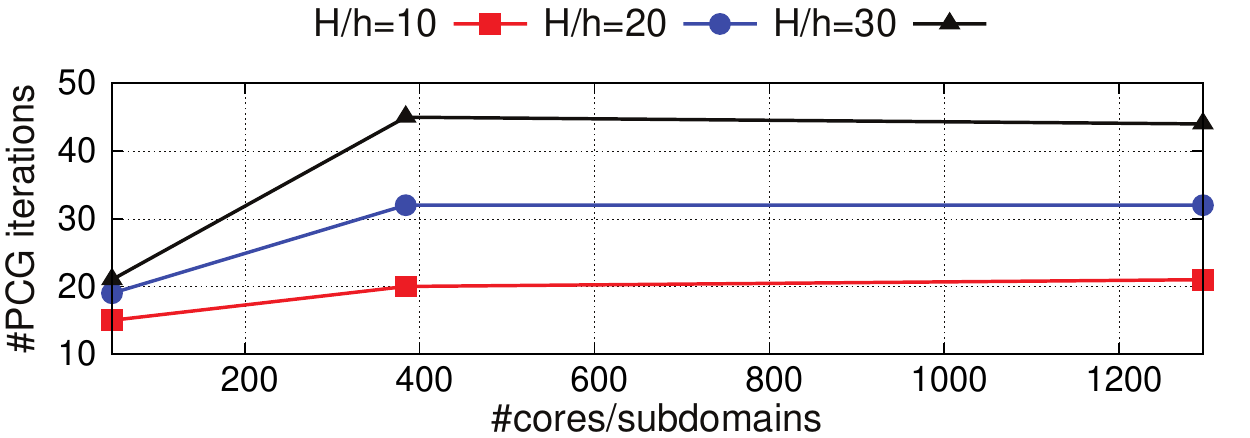}
      \caption{Number of iterations.}
      \label{fig-hts_ifs} 
  \end{subfigure}
  \hspace{0.2cm}
  \begin{subfigure}[t]{0.47\textwidth}
    \centering
    \includegraphics[width=\textwidth]{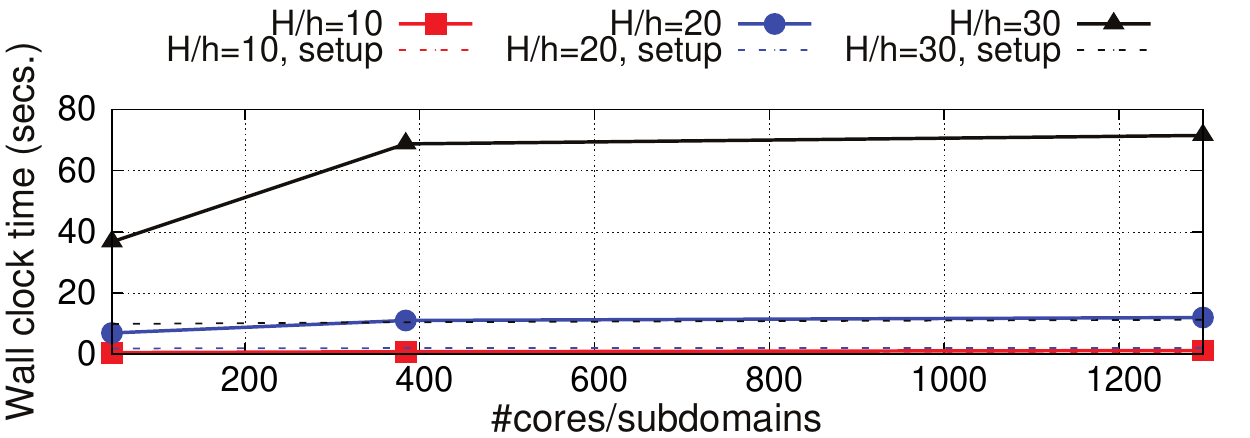}
    \caption{Wall clock time for the total solver and only the set-up phase.}
    \label{fig-hts_tfs}
  \end{subfigure} 
  \caption{Weak scalability for the first linear solver in the \ac{hts} problem with $r=10^2$.}
  \label{fig-hts_fs}
\end{figure}

We first present weak scalability results for the first set-up and solve with the \ac{rpbbddc} preconditioner in \fig{fig-hts_fs}, i.e., the first linearized problem (\eq{eq-linearized_form}) for the first time step. We include results for $\frac{H}{h}=\{10,20,30\}$. As expected, the method shows good weak scalability properties in number of iterations (see \fig{fig-hts_ifs}) and computing times (see \fig{fig-hts_tfs}).

Next, we present average counters for the total number of linear solver applications for the simulation of the whole time interval $[0, 5]$~ms in Table \ref{tab-hts}, for a local problem size of $H/h=10$ and different partitions. The resulting aggregation of cells into subsets based on their physical coefficient $\curlc$ (see \sect{subsec-rpb_imp}) for $t=4$~ms is depicted in \fig{fig-subsets_hts}. We can identify two main regions in the distribution of $\curlc$ (see \fig{fig-alpha_hts}): an inner region that is still not magnetized (i.e., with null resistivity) and a surrounding region, separated by a thin layer. Therefore, the selected value for $r$ allows us to capture the behaviour of the different regions in $\Omega_{\rm hts}$.  
 Out of the results in \tab{tab-hts}, the most salient property is the (asymptotic) scalability in the average number of iterations. Besides, we show how the coarse problem size for the presented cases is only (approximately) doubled regarding to the size that would be obtained with the partition $\partition$ instead of the $\pbpartition$. 

\begin{table}
\centering
\begin{tabular}{ccccc}\hline
\bf{P}      &  \thead{\# Average \\ iters.}  & \thead{Average \\ size($A_c$)}  & \thead{size($A_c$) \\ ratio} &     \\ \hline
\bf{49}     &  16.8     &   196.7     & 1.31$c_0$  \\ 
\bf{385}    &  18.9     &   2424.5    & 2.02$c_0$   \\ 
\bf{1297}   &  21.7     &   7728.7    & 1.91$c_0$   \\  \hline
\end{tabular}
\caption{ Average metrics for the simulation of the time interval $T=[0, 5]$~ms. $c_0$ denotes the number of coarse \acp{dof} of the original, geometrical partition. }
\label{tab-hts}
\end{table} 

\section{Conclusions}\label{sec-conclusions}

In this work, we have proposed an extension of the BDDC preconditioners for arbitrary order curl-conforming spaces that are robust for heterogeneous problems with high contrast of coefficients. 
 The main idea is to enrich the continuity constraints enforced among subdomains (i.e., coarse \acp{dof}) for those which contain high contrast of coefficients. The approach, which is shown to be robust for the grad-conforming case in \cite{BadiaNguyen_physics_based}, makes use of the knowledge about the physical coefficients to define a sub-partition of the original edge \ac{fe}-based definition of coarse objects (edges and faces).
 The motivation for that is the well-known robustness of \ac{dd} methods when there are only jumps of physical coefficients across the interface between subdomains. 
 However, our case is more complex than the one in \cite{BadiaNguyen_physics_based} for Poisson and elasticity problems, since two different coefficients are involved in the time-domain quasi-static approximation to the Maxwell's equations. Our solution is to add a perturbation term to the preconditioner so as we recover a scenario similar to the one in which only one coefficient jumps across the interface.
A relaxed definition of the \ac{pb} subdomains, where we only require that the maximal contrast of the two physical coefficients is smaller than a predefined thresholds, allows one to extend the range of applicability of the preconditioner to truly heterogeneous materials. 
Our preconditioners, which use the crucial change of variables in \cite{toselli_dual-primal_2006} to obtain weakly scalable algorithms for problems in $H$(curl) with few modifications to the standard \ac{bddc} algorithm in \cite{dohrmann_preconditioner_2003}, are empirically shown to be robust with the contrast of coefficients. We would like to remark that our preconditioners maintain the simplicity of the standard \ac{bddc} and do not require to solve any eigenvalue or auxiliary problem.  
    
We devoted a section to describe all the non-trivial implementation issues behind the method based on our experience through the implementation of the preconditioners in \FEMPAR. Its task-overlapping implementation of the \ac{pbbddc} preconditioner allows one to mask the computing times for the coarse problem, as long as they do not exceed local solvers time. With such implementation, we have been able to provide notable weak scalability results in the application of our new preconditioners to a wide range of multi-material and heterogeneous electromagnetics problems, including realistic 3D problems where coefficients can be defined by arbitrary functions, even dependent on the solution itself. In the future, the multilevel extension of the algorithm is expected to push forward the limits of its scalability properties.

\bibliographystyle{my_unsrt}
\bibliography{art031}
\end{document}